\documentclass[graybox,natbib]{svmult}

\usepackage{mathptmx}       
\usepackage{helvet}         
\usepackage{courier}        
\usepackage{type1cm}        
\usepackage{makeidx}         
\usepackage{graphicx}        
\usepackage{multicol}        
\usepackage[bottom]{footmisc}

\usepackage{natbibspacing}

\makeindex             

\begin{document}

\newcommand{\cms}{$\rm cm^{-2}\ $}
\newcommand{\cmq}{$\rm cm^{-3}$}
\newcommand{\kms}{$\rm km\ s^{-1}$}
\newcommand{\ergps}{${\rm erg\ s^{-1}}$}
\newcommand{\lpc}{L_{\rm pc}}
\newcommand{\beq}{\begin{equation}}
\newcommand{\eeq}{\end{equation}}
\def\aa         {{\AA\ }}
\def\stacksymbols #1#2#3#4{\def\theguybelow{#2}
        \def\verticalposition{\lower#3pt}
        \def\spacingwithinsymbol{\baselineskip0pt\lineskip#4pt}
        \mathrel{\mathpalette\intermediary#1}}
\def\intermediary #1#2{\verticalposition\vbox{\spacingwithinsymbol
        \everycr={}\tabskip0pt
        \halign{$\mathsurround0pt#1\hfil##\hfil$\crcr#2\crcr
                \theguybelow\crcr}}}
\def\lta{\stacksymbols{<}{\sim}{2.5}{.2}}
\def\gta{\stacksymbols{>}{\sim}{3}{.5}}
\newcommand{\thin}{\thinspace}
\def\approxprop{\stacksymbols{\propto}{\sim}{3}{.5}}
\newcommand{\chandra}{{\em Chandra}}
\newcommand{\suzaku}{{\em Suzaku}}
\newcommand{\ixo}{\href{http://ixo.gsfc.nasa.gov/}{{\em IXO}}}
\newcommand{\astroh}{\href{http://astro-h.isas.jaxa.jp/}{{\em Astro-H}}}
\newcommand{\hnought}{${\rm H_0}$}
\newcommand{\xmm}{{\em XMM}}
\newcommand{\fbaryon}{${\rm f_{baryon}}$}
\newcommand{\rosat}{{\em ROSAT}}
\newcommand{\asca}{{\em ASCA}}
\newcommand{\rvir}{{\rm r_{vir}}}
\newcommand{\rgs}{RGS}
\newcommand{\hetg}{HETG}
\newcommand{\letg}{LETG}
\newcommand{\nh}{${\rm N_H}$}
\newcommand{\degmark}{$^\circ$}
\newcommand{\xspec}{XSPEC}
\newcommand{\ie}{i.e.}
\newcommand{\eg}{e.g.}
\newcommand{\lcdm}{$\Lambda$CDM}
\newcommand{\lstar}{${\rm L_*}$}
\newcommand{\reff}{${\rm R_e}$}
\newcommand{\msun}{{\rm M_\odot}}
\newcommand{\mvir}{{\rm M_{vir}}}
\newcommand{\lsun}{${\rm L_\odot}$}
\newcommand{\lopt}{${\rm L_{opt}}$}
\newcommand{\arcsec}{$^{\prime \prime}$}
\newcommand{\arcmin}{$^{\prime}$}
\newcommand{\zfe}{${\rm Z_{Fe}}$}
\newcommand{\lx}{${\rm L_{X}}$}
\newcommand{\dtwentyfive}{${\rm D_{25}}$}
\newcommand{\tx}{${\rm T_{X}}$}
\newcommand{\lb}{${\rm L_{B}}$}
\newcommand{\lk}{${\rm L_{K}}$}
\newcommand{\sigmaeight}{$\rm \sigma_8$}
\newcommand\sige{\hbox{{$\sigma_8$}}}
\def\spose#1{\hbox to 0pt{#1\hss}}
\def\ltsim{$\mathrel{\spose{\lower 3pt\hbox{$\sim$}}
        \raise 2.0pt\hbox{$<$}}$\thinspace}
\def\gtsim{$\mathrel{\spose{\lower 3pt\hbox{$\sim$}}
        \raise 2.0pt\hbox{$>$}}$\thinspace}

\newcommand{\mbh}{${\rm M_{BH}}$}
\newcommand{\sigmac}{$\sigma_*$}
\newcommand\logten{\hbox{{$\log_{10}$}}}

\newcommand\aj{{AJ }}%
\newcommand\araa{{ARA\&A }}%
\newcommand\apj{{ApJ }}%
\newcommand\apjl{{ApJ }}%
\newcommand\apjs{{ApJS }}%
\newcommand\ao{{Appl.~Opt. }}%
\newcommand\apss{{Ap\&SS }}%
\newcommand\aap{{A\&A }}%
\newcommand\aapr{{A\&A~Rev. }}%
\newcommand\aaps{{A\&AS }}%
\newcommand\azh{{AZh }}%
\newcommand\baas{{BAAS }}%
\newcommand\jrasc{{JRASC }}%
\newcommand\memras{{MmRAS }}%
\newcommand\mnras{{MNRAS }}%
\newcommand\pra{{Phys.~Rev.~A }}%
\newcommand\prb{{Phys.~Rev.~B }}%
\newcommand\prc{{Phys.~Rev.~C }}%
\newcommand\prd{{Phys.~Rev.~D }}%
\newcommand\pre{{Phys.~Rev.~E }}%
\newcommand\prl{{Phys.~Rev.~Lett. }}%
\newcommand\pasp{{PASP }}%
\newcommand\pasj{{PASJ }}%
\newcommand\qjras{{QJRAS }}%
\newcommand\skytel{{S\&T }}%
\newcommand\solphys{{Sol.~Phys. }}%
\newcommand\sovast{{Soviet~Ast. }}%
\newcommand\ssr{{Space~Sci.~Rev. }}%
\newcommand\zap{{ZAp }}%
\newcommand\nat{{Nature }}%
\newcommand\iaucirc{{IAU~Circ. }}%
\newcommand\aplett{{Astrophys.~Lett. }}%
\newcommand\apspr{{Astrophys.~Space~Phys.~Res. }}%
\newcommand\bain{{Bull.~Astron.~Inst.~Netherlands }}%
\newcommand\fcp{{Fund.~Cosmic~Phys. }}%
\newcommand\gca{{Geochim.~Cosmochim.~Acta }}%
\newcommand\grl{{Geophys.~Res.~Lett. }}%
\newcommand\jcap{{JCAP }}%
\newcommand\jcp{{J.~Chem.~Phys. }}%
\newcommand\jgr{{J.~Geophys.~Res. }}%
\newcommand\jqsrt{{J.~Quant.~Spec.~Radiat.~Transf. }}%
\newcommand\memsai{{Mem.~Soc.~Astron.~Italiana }}%
\newcommand\nphysa{{Nucl.~Phys.~A }}%
\newcommand\physrep{{Phys.~Rep. }}%
\newcommand\physscr{{Phys.~Scr }}%
\newcommand\planss{{Planet.~Space~Sci. }}%
\newcommand\procspie{{Proc.~SPIE }}%
\let\astap=\aap 
\let\apjlett=\apjl 
\let\apjsupp=\apjs 
\let\applopt=\ao

\title*{Dark Matter in Elliptical Galaxies}
\author{David A. Buote and Philip J. Humphrey}
\institute{Department of Physics and Astronomy, 4129 Frederick Reines Hall, University of California at Irvine, Irvine, CA 92697-4575, \email{buote@uci.edu,phumphre@uci.edu} }

%
%
\maketitle

\abstract*{We review X-ray constraints on dark matter in giant elliptical
galaxies $(\rm 10^{12}\, M_{\odot} \lta M_{vir}\lta 10^{13}\,
M_{\odot})$ obtained using the current generation of X-ray satellites,
beginning with an overview of the physics of the hot interstellar
medium and mass modeling methodology. Dark matter is now firmly
established in many galaxies, with inferred NFW concentration
parameters somewhat larger than the mean theoretical relation. X-ray
observations confirm that the total mass profile (baryons+DM) is close
to isothermal $(M\sim r)$, and new evidence suggests a more general
power-law relation for the slope of the total mass profile that varies
with the stellar half-light radius. We also discuss constraints on the
baryon fraction, super-massive black holes, and axial ratio of the
dark matter halo. Finally, we review constraints on non-thermal gas
motions and discuss the accuracy of the hydrostatic equilibrium
approximation in elliptical galaxies.}

\abstract{We review X-ray constraints on dark matter in giant elliptical
galaxies $(\rm 10^{12}\, M_{\odot} \lta M_{vir}\lta 10^{13}\,
M_{\odot})$ obtained using the current generation of X-ray satellites,
beginning with an overview of the physics of the hot interstellar
medium and mass modeling methodology. Dark matter is now firmly
established in many galaxies, with inferred NFW concentration
parameters somewhat larger than the mean theoretical relation. X-ray
observations confirm that the total mass profile (baryons+DM) is close
to isothermal $(M\sim r)$, and new evidence suggests a more general
power-law relation for the slope of the total mass profile that varies
with the stellar half-light radius. We also discuss constraints on the
baryon fraction, super-massive black holes, and axial ratio of the
dark matter halo. Finally, we review constraints on non-thermal gas
motions and discuss the accuracy of the hydrostatic equilibrium
approximation in elliptical galaxies.}

\section{Introduction}
\label{intro}

Cold, non-baryonic dark matter (CDM) is a critical ingredient of the
widely accepted Big Bang cosmological
paradigm~\citep{feng05a,hann06a,eina09a}.  Although the dark matter
(DM) particle has yet to be directly detected in a terrestrial
laboratory~\citep{stef09a,iras09a,ellij10a}, the total amount of DM is
known precisely from observations of the large-scale geometry of the
Universe, in particular the cosmic microwave background radiation,
distant Type Ia supernovas, baryon acoustic oscillations, galaxy
cluster mass functions and baryon fractions. These observations
indicate that DM comprises about 23\% of the energy density budget of
the Universe today compared to about 4\% for baryonic
matter~\citep{pdg08,laha10a}.

While the standard cosmological model describes the large-scale
Universe very well, it has encountered interesting tension with
observations of DM on galaxy scales. Perhaps most notably, beginning
with the seminal studies of rotation curves of dwarf spiral
galaxies~\citep{moor94a,flor94a}, it is now generally recognized that
both small and large disk galaxies have central radial DM profiles
that are flatter than initially predicted by the standard model when
considering only the gravitational influence of the
DM~\citep{gilm08a,prim09a,debl10a}.  Similar cored DM profiles are
obtained by optical studies of elliptical galaxies at the center of
galaxy clusters~\citep{sand02b,kels02a,sand04b,sand08a,newm09a},
though X-ray observations of some clusters find ``cuspy'' total mass
profiles consistent with theory~\citep{arab02a,lewi03a}.  This
``core-cusp'' problem has generated great interest as it may yield
information on the nature of the DM that is apparently inaccessible to
observations on larger scales. An influential early suggestion to
reconcile observation and theory is that the DM particles are
self-interacting~\citep{sper00,koch00a,firm01a,ahn05a}, though
observations apparently do not currently favor that
possibility~\citep{arab02a,lewi03a,rand08a,kuzi10a}.

Alternatively, the core-cusp problem may reveal clues to the influence
of baryons on galaxy formation which should be most important near the
centers of galaxies. In the standard model slightly over-dense regions
in a nearly uniform early DM distribution acted as seeds for future
galaxies and clusters. Baryonic material subsequently fell into the
potential wells established by the DM.  The interplay of the baryons
and DM through, e.g., adiabatic contraction and dynamical friction,
has likely altered the spatial distributions of both quantities during
galaxy formation and
evolution~\citep{blum84a,elza04a,gned07a,roma08a,delp09a,duff10a,abad10a}.
Hence, the DM profiles on galaxy scales can serve as important
laboratories for studying galaxy formation and the nature of the DM
particle.

Like all the different types of galaxies, giant elliptical galaxies
have their own special properties that make them deserving of
dedicated study. Unlike disk galaxies or the smallest dwarf
elliptical galaxies, giant elliptical galaxies are sufficiently
massive that multiple powerful techniques to measure their mass
profiles are available for many systems. In particular, only for giant
ellipticals can gravitational lensing and X-rays be used to complement
the information available from stellar dynamics, the latter of which
is also available for smaller galaxies. Also, in contrast to the
smallest galaxies, every giant elliptical is thought to host a
super-massive black hole (SMBH) at its center, the formation of which
is apparently intertwined with the host
galaxy~\citep{ferr00a,gebh00a}. It is also becoming increasingly
recognized that galaxies acquire their gas in two modes that are
distinguished by whether or not the in-falling gas is heated
\citep{birn03a,kere05a}. Both modes can operate in the transition mass
regime of giant elliptical galaxies
($10^{12}-10^{13}M_{\odot}$)~\citep{birn07a}.

One of the advantages of using X-ray observations to probe DM and
SMBHs in elliptical galaxies is that the hot X-ray--emitting
interstellar medium (ISM) fills the three-dimensional galactic
potential well, thus providing a continuous tracer from the galactic
nucleus out to well past the optical half-light radius where stellar
dynamical studies become very challenging.  In order to translate an
X-ray observation of the hot ISM into a gravitating mass profile and
DM measurement, it is required that hydrostatic equilibrium holds to a
good approximation throughout the region of interest; i.e., that the
thermal gas pressure of the ISM balances the weight of the gas. With
X-ray observations providing increasing evidence of morphologically
disturbed elliptical galaxies, it is timely to review the accuracy of
the approximation of hydrostatic equilibrium in X-ray determinations
of their DM profiles.  Understanding and quantifying deviations from
hydrostatic equilibrium not only will lead to more accurate X-ray mass
determinations, but any measured non-hydrostatic gas motions will
provide clues to elliptical galaxy
formation~\citep{math03a,birn03a,kere05a,birn07a,deke08a,brig09a}.

Since giant elliptical galaxies tend to be located in dense
environments, it can be difficult to disentangle any DM associated
with the elliptical galaxy from its parent halo. Naturally, most of
the attention of X-ray studies has been devoted to those systems with
the largest X-ray fluxes which has lead to an emphasis on elliptical
galaxies located at the centers of the most massive galaxy groups and
clusters.  While all elliptical galaxies are of interest for DM
studies, in this review we address galaxy-sized halos by focusing on
the lower mass regime that is still accessible to detailed X-ray study
($\rm 10^{12}\, \msun \lta \mvir \lta \rm  10^{13}\,
\msun$) We restrict consideration to objects where a single giant
elliptical galaxy dominates the stellar light, which at the upper end
of the mass range under consideration are often classified as fossil
groups~\citep{ponm94,vikh99a,jone03a} which are thought to be highly
evolved, relaxed systems~\citep{dong05a,dari07a,milo06a}.

In the present review we focus on the constraints on DM in elliptical
galaxies obtained using the current generation of X-ray observatories
-- \chandra, \xmm, and \suzaku; constraints from the previous
generation of X-ray satellites are reviewed by, e.g.,
\cite{fabb89,buot98a,math03a}. While we take care to mention relevant
results obtained from other techniques, our review is not intended to
be complete from the multi-wavelength perspective. The interested
reader is urged to consult the reviews by~\citep{gerh06a,gerh10a}
and~\citep{elli10a,treu10a} for discussions of the constraints on DM
in elliptical galaxies obtained by stellar dynamics and gravitational
lensing.

\section{Hydrostatic Models of the Hot ISM}
\label{he}

\subsection{Preliminaries}
\label{pre}

The hot ISM of an elliptical galaxy $(T\sim 10^7\,{\rm K}\sim 0.9\,
{\rm keV}/k_B)$ has properties that are extremely similar to the
(hotter) intracluster medium (ICM) of a massive galaxy
cluster~\citep{sara86a,fabi90a,sara92a}. While the detailed radiation
spectrum is described well by a collisionally ionized {\it plasma} of
electrons and ions obeying the coronal approximation~\citep{mewe99a},
the macro-structure of the hot ISM in an elliptical galaxy is for many
purposes also described well by a simple {\it monatomic ideal gas}. We
highlight four key features of the ``hot gas'' relevant for
constructing hydrostatic models to measure the DM content in
elliptical galaxies.

\paragraph{\bf Collisional Fluid} 

In a fully ionized plasma, the equipartition timescale for electrons
considering only their mutual Coulomb interactions is~\citep{spit62a},
\begin{eqnarray}
\tau_{e-e} & = &\frac{3m_e^{1/2}\left(k_BT_e\right)^{3/2}}{4\pi^{1/2}n_e
e^4\ln\, \Lambda_{\rm coul}}\nonumber\\
& = & 1.1\times 10^4\, \, \left(\frac{T_e}{10^7\, \rm K}\right)^{3/2}
\left(\frac{n_e}{10^{-3}\, \rm cm^{-3}}\right)^{-1}\, {\rm yr}, \label{eqn.tee}
\end{eqnarray}
where $m_e$ is the electron mass, $e$ is the electric charge, $k_B$ is
Boltzmann's constant, $n_e$ is the electron number density, and $T_e$
is the electron temperature. For $T_e > 4\times 10^5$~K the Coulomb
logarithm of an electron-proton plasma is~\citep{spit62a},
\begin{equation}
\ln\, \Lambda_{\rm coul}  =  35.4 + \ln\left[\left(\frac{T}{10^7\, \rm K}\right)
\left(\frac{n_e}{10^{-3}\, \rm cm^{-3}}\right)^{-1/2}\right].
\end{equation}
The mean free path of the electrons for Coulomb collisions is ${\rm
v}_e\tau_{e-e}$~\citep{spit56a,cowi77a,dopi03a}, where ${\rm v}_e=\sqrt{3k_BT_e/m_e}$
is the rms thermal velocity of the electrons, so that,
\begin{eqnarray}
{\rm Coulomb\,\, mean\,\, free\,\, path} & = & \frac{3^{1/2}\left(k_BT_e\right)^2}{4\pi^{1/2}n_e
e^4\ln\, \Lambda_{\rm coul}}\nonumber\\
& = & 243\,\, \left(\frac{T}{10^7\, \rm K}\right)^2
\left(\frac{n_e}{10^{-3}\, \rm cm^{-3}}\right)^{-1}\, {\rm pc}.
\end{eqnarray}
Because $m_e$ does not enter the above equation this expression also
represents the mean free path for the protons (for $n_e=n_p$ and
$T_e=T_p$). The electron and proton mean free paths are generally much
smaller than the length scales of interest for DM studies indicating
that the plasma may be treated as a collisional fluid. Still, in the
outer regions near the virial radius, where $n_e$ is small, and
therefore the Coulomb mean free path is large, it is possible the
fluid approximation considering only Coulomb collisions may not always
be justified. But the hot ISM of elliptical galaxies may typically
possess tangled $\sim 1\mu\, \rm G$ magnetic
fields~\citep{vall04a,math97a}. If so, the electron gyroradius,
$r_g={\rm v_e}/\omega$, for angular gyrofrequency $\omega=eB/m_ec$,
\begin{equation}
r_g = 3.9\times 10^{-11}\, Z^{-1}\left(\frac{T}{10^7\, \rm
K}\right)^{1/2}\left(\frac{m}{m_e}\right)^{1/2} \left(\frac{B}{1\, \mu
G}\right)^{-1}\, \rm pc,
\end{equation}
is vastly smaller than the relevant length scales for DM studies. (The
same is true for the proton gyroradius obtained by setting $m=m_p$.)
Consequently, if the hot ISM contains (plausible) weak magnetic
fields, then it is expected to behave as a collisional fluid on all
length scales of interest for DM studies, allowing an isotropic
pressure and temperature to be defined for any fluid element.

\paragraph{\bf Local Maxwellian Velocity Distribution}  

In any fluid element in the hot ISM Coulomb collisions between
electrons establish a Maxwellian velocity distribution with kinetic
temperature $T_e$ on the time scale given by equation
(\ref{eqn.tee}). This is much shorter than other time scales
associated with the hot ISM in a relaxed elliptical galaxy. In general
the ions can reach equilibrium with a kinetic temperature different
from that of the electrons. But electron-ion collisions will establish
equipartition on a time scale $\approx (m_p/m_e)\tau_{e-e}\approx
1870\tau_{e-e}\approx 10^7$~yr for the conditions typical of the hot
ISM~\cite{spit62a}. Only for galaxies currently undergoing strong
dynamical disturbances, particularly in the central regions of some
galaxies with AGN, might be evolving dynamically on a comparable time
scale so that the electron and ion temperatures could be different. In
relaxed elliptical galaxies, however, the hot ISM is expected to
behave locally as a Maxwellian velocity distribution with a single
kinetic temperature for the electrons and the ions. Therefore, the
thermal pressure of the hot ISM is characterized by that of a
monatomic ideal gas, $\rho_{\rm gas}k_BT/\mu m_{\rm a}$, where $\mu$
is the mean atomic weight of the gas and $m_{\rm a}$ is the atomic
mass unit. For a fully ionized gas of solar abundances, $\mu = 0.62$.

\paragraph{\bf Thermal Pressure Dominates Magnetic Pressure}

Observations of radio halos and relics in several massive galaxy
clusters indicate that the ICM is magnetized with a field strength of
$1-10\,\mu\rm G$~\citep{govo04b}. Unfortunately, elliptical galaxies
are not observed to possess the large-scale diffuse radio emission
observed in some massive clusters, and thus knowledge of their
magnetic fields is derived from systems with embedded radio jets where
the inferred $\sim\mu\rm G$ fields~\citep{vall04a} may not be
representative of the typical hot ISM. For fields of this magnitude,
with $P_{\rm mag} = B^2/8\pi$ and $P_{\rm therm} = n_{\rm gas}k_BT
\approx 1.9n_ek_BT$,
\begin{equation}
\frac{P_{mag}}{P_{therm}} = 0.015\,  \left(\frac{B}{1\, \mu
G}\right)^2 \left(\frac{T}{10^7\, \rm K}\right)^{-1}
\left(\frac{n_e}{10^{-3}\, \rm cm^{-3}}\right)^{-1},
\end{equation}
which implies a minor, though very uncertain, contribution of magnetic
fields to the pressure support of the hot ISM. Many theoretical models
of inflowing gas in elliptical galaxies predict that small seed
magnetic fields are amplified possibly to this level or larger
~\citep[][and references therein]{math03a}, although the small gas
fractions observed at the centers of groups~\citep{gast07b,sun09a}
likely indicates that any hypothetical concentration magnetic fields,
e.g., arising from the amplification owing to inflowing gas just
described, has been dispersed by feedback.  X-ray studies of DM in
elliptical galaxies generally neglect the magnetic pressure, an
assumption that can be tested indirectly by (1) comparing observed
stellar mass-to-light ratios to theoretical stellar population
synthesis models and (2) comparing mass profiles obtained by X-ray
methods with those obtained from stellar dynamics and gravitational
lensing (\S
\ref{acc}).

\paragraph{\bf Thermal Motions Dominate the Hot ISM in Relaxed Galaxies}

When the gravitational potential evolves on a time scale longer than
the dynamical time, approximately given by the crossing time of the
stars,
\begin{equation}
t_{\rm cross} = \frac{d}{\sigma} = 3.9\times 10^7\,
\left(\frac{d}{10\,\, \rm kpc}\right)\left(\frac{\sigma}{250\,\, \rm
km\,\, s^{-1}}\right)^{-1}\,\, \rm yr, \label{eqn.cross}
\end{equation}
where $d$ is the length scale of the region under consideration and
$\sigma$ is the stellar velocity dispersion within that region, a
time-independent description of the gravitational potential is
appropriate. This quasi-static description of the gravitational
potential is valid regardless of the dynamical state of the hot ISM,
because the gas contributes negligibly to the total mass
(gas+stars+DM) over most of the radial range within the virial radius
of an elliptical galaxy.  Radiative cooling of the hot gas acting
alone cannot drive evolution on time scales smaller than the cooling
time ($\approx 5\times 10^6/n_e$~yr for $T=10^7$~K and solar
abundances, or $5\times 10^9$~yr for $n_e=10^{-3}$~cm$^{-3}$) which is
generally much longer than $t_{\rm cross}$. But rapid evolution on a
time scale $\lta t_{\rm cross}$ is expected to prevail in the central
regions of galaxies that have experienced recent, strong AGN feedback
presumably generated in response to gas cooling. For our purposes, a
{\it relaxed} elliptical galaxy is one in which the structure of the
hot ISM is also evolving quasi-statically on a time scale longer than
$t_{\rm cross}$ described by a state of hydrostatic equilibrium, which
applies provided any non-thermal gas motions (i.e., motions that are
not associated with the random velocities that define the thermal gas
pressure) are negligible compared to the sound speed,
\begin{equation}
c_s = \left(\frac{\gamma P_{\rm gas}}{\rho_{\rm gas}}\right)^{1/2} =
473\left(\frac{T}{10^7\, \rm K}\right)^{1/2}\,\, \rm km\,\, s^{-1},
\end{equation}
where $\gamma = 5/3$ is the adiabatic index. This implies a sound
crossing time similar to the stellar crossing time in equation
(\ref{eqn.cross}). Turbulence is expected to be introduced into the
hot ISM from merging and intermittent AGN feedback. Interesting limits
on turbulent motions have been obtained via constraints on resonance
scattering in the center ($\lta 1$~kpc) of the elliptical galaxy
NGC~4636~\citep{wern09a} and from Doppler broadening of the X-ray
emission lines in a small number of elliptical galaxies (NGC~533,
NGC~1399, NGC~4261, NGC~5044) from the sample of~\cite{sand10a}
suggesting that any turbulent motions present in the central regions
of those systems are highly subsonic.  While not expected to dominate,
turbulent velocities may not be negligible since cosmological
hydrodynamical simulations typically find that turbulence contributes
up to $\sim 20\%$ of the pressure support of the ICM in relaxed galaxy
clusters~\cite{tsai94a,evra96a,naga07a,piff08a,fang09a}.  The
irregular, subsonic H$\alpha$ velocities observed at the centers of
some elliptical galaxies also suggest turbulent velocities in the hot
gas~\cite{caon00a}.  It is unlikely that rotation dominates the ISM
dynamics on large scales in elliptical galaxies given the lack of
flattened disks observed~\citep{hanl00a}, though modest rotational
spin-up might occur in the innermost $(\lta 1\,\,\rm kpc)$ regions of
some galaxies~\cite{brig09a}. The expected inflow velocities arising
from radiative cooling are negligible~\citep{math03a}, and a recent
study of galaxy-scale halos using a cosmological hydrodynamical
simulation concludes that their hot ISM is generally
quasi-hydrostatic~\citep{crai10a,crai10b}.

\subsection{The Equation of Hydrostatic Equilibrium: Usage Guidelines}

For the relaxed elliptical galaxies under consideration, hydrostatic
equilibrium of a fluid element of hot ISM is expressed as the balance
between forces per unit volume of gravity and gas pressure,
\begin{equation}
\nabla P_{\rm gas} = -\rho_{\rm gas}\nabla\Phi, \label{eqn.he}
\end{equation}
where $\Phi$ is the gravitational potential, $P_{\rm gas}$ is the
pressure and $\rho_{\rm gas}$ the density of the ISM.  If non-thermal
effects are significant that can be represented by a pressure ($P_{\rm
nt}$), such as for isotropic random turbulence or a magnetic field,
then $P_{\rm gas} = P_{\rm t} + P_{\rm nt}$, where $P_{\rm t}$ is the
thermal pressure. For the case of solid-body rotation, $\Phi$ can be
replaced by an effective potential. For general rotation (and other
ordered gas motions) the convective derivative term
$(\vec{v}\cdot\nabla)\vec{v}$ in the Euler equation must be added to
the left-hand side of equation (\ref{eqn.he}). As remarked previously,
the hydrostatic approximation is appropriate provided the non-thermal
gas motions are subsonic.

In order to measure directly the magnitude of any non-thermal motions
present in the hot ISM, it is necessary to measure precisely the
Doppler shifts and broadening of emission lines which is beyond the
capability of the current X-ray satellites -- except possibly for the
innermost regions of some galaxies noted above. Consequently, indirect
methods are required to assess whether a galaxy is suitably relaxed
for the hydrostatic approximation. Assessment of the morphology of the
X-ray image is a convenient first step to locate candidate relaxed
galaxies for the following reason. Because the total underlying
stellar and DM distribution is expected to have a shape close to that
of an ellipsoid, the gravitational potential is also expected to be
approximately ellipsoidal, though rounder than the generating mass
distribution. Since in hydrostatic equilibrium the X-ray emissivity
traces the same three-dimensional shape as the potential for any
temperature profile \citep[][also see \S
\ref{nonsph}]{buot94,buot96a,buot98a}, galaxies that exhibit
irregular, non-ellipsoidal features in their X-ray images might not be
suitably relaxed for hydrostatic study.  The existence of such
asymmetrical features (e.g., substructure, cold fronts) does not
guarantee large departures from hydrostatic equilibrium, as
demonstrated by cosmological hydrodynamical simulations of
clusters~\citep{tsai94a,evra96a,naga07a,piff08a,fang09a}, especially
when the features are sufficiently localized so that they can be
easily excluded from study~\citep{buot95a,naga07a}.

Another important consideration in hydrostatic analysis is to insure
that the X-ray emission that is attributed to hot ISM is not, in fact,
heavily contaminated by unresolved discrete sources.  A large number
of the brightest discrete sources (mostly low-mass X-ray binaries --
LMXBs) can be detected and excluded in nearby galaxies using the high
spatial resolution of \chandra~\citep[see][and references
therein]{hump08b}; the lower resolution of \xmm\ leads to far fewer
detections, and only a few sources (if any) can be detected with
\suzaku. Because the combined spectrum of unresolved LMXBs is spectrally
harder than the hot ISM, the spectral signature of the hot ISM can be
reliably extracted from the data provided it is not overwhelmed by
the unresolved source component. The situation is more problematic in
X-ray faint galaxies (having low ratios of X--ray-to-optical
luminosity) like NGC~3379, where hydrostatic mass estimates have been
attempted~\cite{fuka06a,pell06a} but the soft, diffuse X-ray emission
attributed to hot ISM most likely originates almost entirely from
cataclysmic variable stars and stellar coronae~\citep{revn08a,trin08a}.

Once a suitably relaxed galaxy with sufficient emission from hot ISM
has been identified, the accuracy of the hydrostatic equilibrium
approximation can be assessed indirectly via consistency
checks. First, one may examine if any well-motivated hydrostatic model
is able to provide an acceptable fit to the observed density and
temperature profiles of the hot ISM. If an acceptable fit is obtained,
the data are consistent with, though do not necessarily require,
hydrostatic equilibrium. A poor fit -- particularly one with sharp
discontinuities in the density and temperature profiles -- provides
strong evidence for significant violations of the hydrostatic
equilibrium approximation \footnote{Still, even in such cases the hydrostatic
approximation might be useful. The core of M~87 -- a system more
massive than we consider in this review -- possesses various spatial
and spectral~\cite{form07a,mill10a} irregularities that are very likely
associated with sizable non-hydrostatic gas motions. Yet
\cite{chur08a} found that the gas is close to hydrostatic away from
the regions of most significant disturbance.}.  Second, the
gravitating mass profile inferred from the X-ray data can be compared
to independent measurements using other techniques having different
assumptions (stellar dynamics, gravitational lensing). When results
obtained from different methods agree, it provides support for the
underlying assumptions of each method and implies that systematic
errors are well understood (see \S \ref{acc}).

Hence, given the limitations of current X-ray data, a sensible
approach to apply and assess hydrostatic equilibrium is the following:

\paragraph{\bf Guidelines for Applying Hydrostatic Equilibrium in
Elliptical Galaxies:}
\begin{itemize}

\item Select systems which tend to have a regular,
approximately cicular or elliptical X-ray image morphology. If the
image has asymmetric features, they should be preferably confined to
spatial regions that are small compared to the region of interest;
e.g., the inner region showing evidence of possible AGN feedback.

\item Unresolved discrete sources should not
dominate the X-ray emission. As a rough guide, at least half the X-ray
luminosity within the optical half-light radius should originate from
hot gas.

\item Determine whether a hydrostatic model is able to provide an
acceptable fit to the density and temperature profiles of the hot ISM
over the region of interest.

\item When possible, compare the mass profile obtained via the
hydrostatic approximation with that obtained by an independent method;
e.g., stellar dynamics, gravitational lensing. The hydrostatic
equilibrium approximation is judged to be useful if it provides DM
measurements of comparable (or better) quality than other techniques
-- or provides any information when no other technique is available.

\end{itemize}

\begin{figure*}[t]
\centerline{\includegraphics[scale=0.59,angle=0]{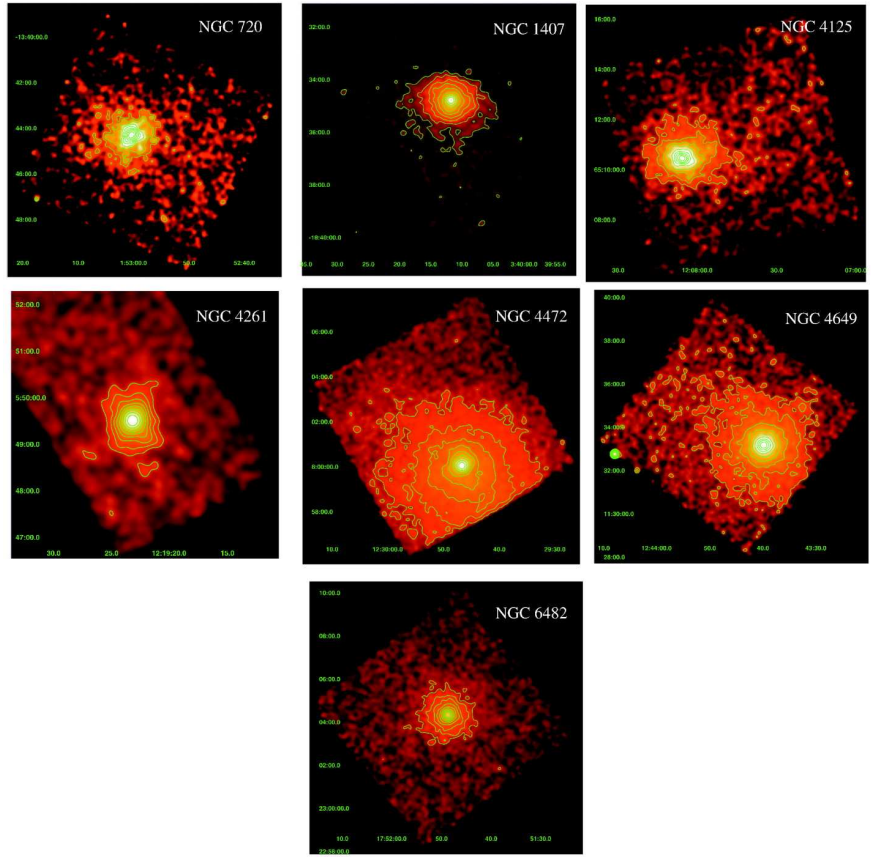}}
\caption{\label{fig.h06}
A sample of seven nearby elliptical galaxies that are suitable for
hydrostatic analysis, with each possessing regular morphologies as
revealed by their \chandra\ images~\cite{hump06b}.}
\end{figure*}

\begin{figure*}[t]
\centerline{\includegraphics[scale=0.59,angle=0]{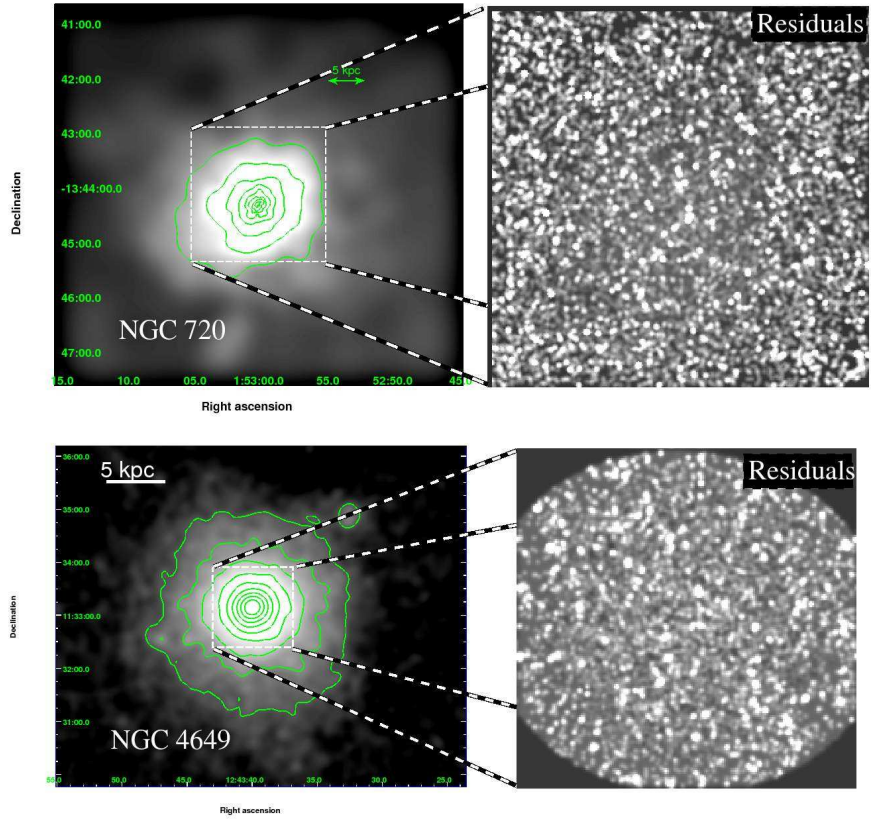}}
\caption{\label{fig.resid}
Residual significance images of the central regions of NGC~720 (top,
adapted from \cite{hump11a}) and NGC~4649 (bottom), indicating
deviations from a smooth model fit to the X-ray
isophotes~\citep{hump08a,hump11a}. Both galaxies appear to be relaxed
and symmetrical at all scales accessible to the \chandra\ images.}
\end{figure*}

The disturbed X-ray image of M~84 is a spectacular
example~\citep{cava10a} of the profound effect that AGN feedback can
have on the hot ISM and is not a galaxy where hydrostatic equilibrium
is expected to apply very accurately. But not all elliptical galaxies
display such strong, large-scale irregularities in their X-ray images.
In Figure~\ref{fig.h06} we display several elliptical galaxies with
mostly regular images suitable for hydrostatic analysis. Two of the
them, NGC~720 and NGC~4649, have very regular images deep into their
cores (Figure ~\ref{fig.resid}). Another two, NGC~4472 and NGC~4261,
possess low-level image fluctuations consistent with cavities
associated with radio jets that do not greatly interfere with
hydrostatic mass analysis.  For several of these galaxies, radial
entropy profiles have been computed~\citep{hump08a,hump09c,hump11a}
and found to increase monotonically with radius as also found in
groups and clusters~\citep{prat10a}, indicating that the hot ISM is
stable to convection and consistent with approximate hydrostatic
equilibrium.

\subsection{On the Incidence of Relaxed Elliptical Galaxies}

In a controversial study, Diehl \& Statler~\citep{dieh07a}
investigated the X-ray and optical isophotal ellipticities in a
heterogeneous sample of early-type galaxies to test whether they are
consistent with hot ISM obeying hydrostatic equilibrium. The expected
correlation between the ellipticities assumes that both hydrostatic
equilibrium holds exactly and mass follows (optical) light, so that
the stars define the shape of the gravitational potential and thus the
shape of the X-ray emission. In an attempt to mitigate the impact of
any DM halo, Diehl \& Statler restricted their analysis of each galaxy
to within a small radius -- 60\%-90\% of the optical half-light
radius.  When the expected correlation between the X-ray and optical
isophotal ellipticities was not found, they concluded that hydrostatic
equilibrium is not ubiquitous and, therefore, should never be assumed
(even approximately) in mass analysis.  They speculate that assuming
hydrostatic equilibrium will always result in an error typically as
large as the value of the mass measured; i.e., relaxed elliptical
galaxies do not exist. While we agree (as detailed in the previous
section) that care should be exercised in the application of
hydrostatic equilibrium to the study of elliptical galaxies, we
strongly disagree with their general conclusions about the accuracy of
X-ray mass measurements.

The observed lack of a correlation does not allow conclusions to be
drawn with certainty for an individual object; it only indicates that
hydrostatic equilibrium is unlikely to hold perfectly in a galaxy
randomly drawn from their sample, provided it is chosen without any
consideration of its morphology.  As noted above, hydrostatic
equilibrium as expressed by equation~(\ref{eqn.he}) is an
approximation and is not expected to hold exactly given especially the
impact of turbulence on the hot ISM induced by past mergers and AGN
feedback. In other words, no elliptical galaxy is expected to be
perfectly relaxed. The objects displayed in Figures~\ref{fig.h06}
and~\ref{fig.resid} are among the most relaxed elliptical galaxies
possessing high-quality X-ray data.  But Diehl \& Statler's sample
also contained galaxies with pronounced large-scale asymmetries that
we would not recommend the routine application of hydrostatic mass
methods following the guidelines discussed previously (e.g.,
NGC~4636:~\cite{jone02a} and M~84:~\cite{cava10a}), and since they
focused on the smallest scales (where X-ray point-source removal is
most challenging and where AGN-driven disturbances are most serious),
in contrast to the larger scales that are most important for DM
analysis, the implications of their study for DM measurements of
morphologically relaxed systems are dubious and provide no basis for
claiming typical mass errors $\gta 100\%$ on all scales in such
systems.

An example of how conclusions regarding the center of a galaxy need
not apply on larger scales is provided by NGC~5044. In this system
Diehl \& Statler measured an X-ray ellipticity ($0.41\pm 0.08$)
greatly exceeding the optical value ($0.07\pm 0.01$), which cannot be
reconciled easily with hydrostatic equilibrium (unless the gas is also
rapidly rotating).  But the X-ray ellipticity measurement is confined
to the region within the optical half-light radius ($<4$~kpc)
containing obvious disturbances presumed to be associated with AGN
feedback~\citep{buot03a,gast09a,davi09a}.  (We note, however, that
there is recent evidence that turbulent motions of the hot ISM are
negligible compared to the sound speed in the center of
NGC~5044~\citep{sand10a}.)  The X-ray emission can be traced out at
least to several hundred kpc, about one-third of the virial
radius~\citep{davi94,buot04b}.  Outside of the very central region
discussed by Diehl \& Statler, the X-ray isophotes are largely
symmetric and nearly circular, and acceptable fits with hydrostatic
models are obtained~\citep{gast07b}.

Now turning our attention to the central regions within the optical
half-light radius relevant to the study of SMBHs (\S \ref{smbh}), we
remark that additional factors can introduce scatter into the
correlation between the isophotal ellipticities generated by the stars
and hot ISM. As noted above, Diehl \& Statler assume that the mass
traces the same shape as the optical light, whereas both X-ray and
optical observations suggest DM distributed differently from the stars
typically contributes 20\%-30\% within the half-light radius in giant
elliptical galaxies (see \S \ref{rad}). Furthermore, their novel
technique to create X-ray images of diffuse hot ISM strictly requires
there are no spatial variations in the spectral properties; i.e.,
every galaxy is assumed to be isothermal with constant metallicity
throughout, in conflict with observations of many galaxies possessing
high-quality X-ray
data~\citep{buot02a,buot03b,kim04a,hump06a,fuka06a,mats07a,rasm07a,komi09a}.
Finally, ISM rotation might be important in some galaxies and yet have
little relation to stellar rotation because slowly rotating inflowing
gas at large radius might be spun up to dynamically relevant
velocities if angular momentum is conserved. But rotational spin-up is
not guaranteed for every galaxy, and instead the gas can be outflowing
gently in a subsonic wind greatly reducing the impact of any rotation
initially provided by the stars -- see discussion in~\citep{math03a}.

Before leaving this topic, we caution that a statistical appraisal of
the properties of the hot ISM in any sample of elliptical galaxies
should consider the sample selection criteria. Our impression of the
X-ray properties of elliptical galaxies is obtained within the context
of what are essentially optically selected samples.  But for the study
of ISM properties, it is important to select the sample according to
the ISM luminosity and not the total X-ray luminosity which also
contains the emission from discrete sources associated with the
stellar population.  Diehl \& Statler follow the common approach of
studying a ``heterogeneous'' sample of elliptical galaxies, which
essentially amounts to collecting all objects in the archives with
data of sufficient quality.  This procedure resulted in a sample
containing some well known objects with extended, X-ray luminous ISM
(e.g., NGC~4472) and some where virtually all the X-ray emission can
be attributed to discrete sources (e.g., NGC~3379 --
\cite{revn08a,trin08a}).

It is instructive, therefore, to consider a sub-sample extracted from
their sample of 54 galaxies that is selected based on the X-ray
luminosity of the diffuse gas. Using the tabulated diffuse gas
luminosities published by Diehl \& Statler, we construct such a
sub-sample by excluding massive group/cluster systems (gas $L_{\rm x} >
3\times 10^{42}$~\ergps) and gas-poor galaxies ($2\sigma$ lower limit
on gas $L_{\rm x} < 1\times 10^{40}$~\ergps) resulting in a sample of
26 galaxies, give or take a few depending on how we interpret the
quoted lower or upper limits for some systems. This sub-sample includes
the seven galaxies shown in Figure~\ref{fig.h06}, as well as several
other systems with generally regular X-ray images and where
hydrostatic equilibrium previously has been profitably employed --
NGC~533, NGC~1399, NGC~1404, NGC~3923, NGC~4555, NGC~5044. The sample
also contains well-known disturbed systems such as M~84, NGC~4636, and
NGC~5846, which are expected to deviate from hydrostatic
equilibrium. In sum, an ISM-selected sub-sample culled from Diehl \&
Statler's heterogeneous sample indicates that approximately 50\% of
the elliptical galaxies are sufficiently relaxed for hydrostatic
analysis.

\subsection{Methods}

The variety of techniques that have been developed to apply
hydrostatic equilibrium to the study of the mass distributions in
elliptical galaxies (as well as galaxy clusters) can be divided into
two broad classes that will be denoted here as {\it smoothed
inversion} and {\it forward-fitting}.  Whereas the forward-fitting
approach assumes parameterized models for the mass profile and one
thermodynamical variable of the gas (e.g., temperature profile),
smoothed inversion instead introduces one or more arbitrary parameters
to guarantee a smooth (physical) mass profile.  The decision of
whether or not to adopt a particular approach depends on several
factors which we discuss below. Generally, the smoothed inversion
approach is preferred if one is primarily interested in measuring the
value of the enclosed mass within some radius without making any
reference to an assumed (parameterized) mass model. If instead one
desires to test the viability of an assumed mass model (e.g.,
power-law, NFW) and to measure its parameters for comparison to other
galaxies and theory, the forward-fitting approach is
preferred. Whenever possible both approaches should be employed to
assess the magnitude of any biases.

We note that in the following it is generally assumed that the gas
density and temperature profiles (projected or deprojected) of the hot
gas have been already obtained via an analysis of the spatially
resolved X-ray spectra. The hydrostatic models then use these measured
profiles to constrain the mass distribution.  This decoupling of the
spectral fitting and mass modeling, though convenient and generally
less computationally expensive, is not necessary. The formalism of
both of these approaches is reviewed in Appendix B of~\citep{gast07b}.

In the following subsections, we outline in more detail how the two
broad model classes are implemented.

\subsubsection{Smoothed Inversion}

\paragraph{\bf Pros:}
\begin{itemize}
\item Mass measured without reference to an input parameterized
model.
\item Does not require an input parameterized model for the
gas temperature, density, pressure, or entropy.
\end{itemize}

\paragraph{\bf Cons:}
\begin{itemize}
\item Deprojection requires a high degree of symmetry (usually spherical).
\item Cannot self-consistently account for the projection of galaxy emission from any
radii outside of the observed data range or, generally, extrapolate
the hydrostatic model outside the data range.
\item Arbitrary smoothing required to obtain a physical (monotonically
increasing) mass profile. The amount of smoothing increases for lower
data quality.
\end{itemize}

The aim of the smoothed inversion approach is to measure the mass
profile without reference to an assumed (parameterized) model for the
mass distribution or any quantity associated with the hot ISM (gas
temperature, density, pressure, or entropy).  For this to be achieved,
the three-dimensional distribution of the hot ISM must be obtained by
non-parametrically deprojecting the data on the sky.  This is only
possible if the ISM has a high degree of symmetry, and therefore
spherically symmetric deprojection using the well-known ``onion peel''
method originally developed for clusters is almost always
adopted~\citep{deproj,kris83}. Since the onion peeling starts from the
outer edge of the galaxy and works its way inward, such deprojection
formally requires that the data span the entire size of the galaxy. As
this is not generally the case, ad hoc prescriptions are usually
employed to estimate the projection of gas from radii larger than the
extent of the available data~\citep{nuls95a,buot00c}.

Because no global mass model is assumed, the value of the mass
determined within a given shell can be negative due to noise in the
data (amplified by deprojection). To circumvent this problem so as to
achieve a mass profile that increases monotonically with radius, the
inversion procedure must smooth the gas quantities (e.g., temperature
and density) and the mass profile itself. The amount and type of
smoothing is arbitrary, and typically as the data quality decreases
the amount of smoothing required to achieve interesting constraints
generally increases.  As discussed below, types of smoothing include
(1) the use of ad hoc smooth parameterized models for the gas density
and temperature, (2) assuming the ISM is constant in spherical shells
so that the widths of the shells set the smoothing scales, and (3) one
or more parameters in a Bayesian prior. While the smoothing can be
well-motivated, it necessarily introduces additional assumptions,
usually expressed by additional parameters that must be determined
empirically by the data.  Consequently, we believe that the terms
``non-parametric'' or ``model-independent'' sometimes used to describe
smoothed inversion methods are misleading and should be avoided.

In the context of DM studies, smoothed inversion is best suited to
providing a measurement of the enclosed mass within a particular
radius without assuming a specific parameterized mass model; e.g.,
measuring $M_{500}$ for the purpose of establishing the mass function
of elliptical galaxies.  But if one also desires to fit and measure
the parameters of a mass model (e.g., global power-law slope, NFW
concentration parameter), the ad hoc smoothing adds a barrier between
the data and model which is not present in the forward-fitting
approach that is more naturally suited for this problem.

\paragraph{\bf Traditional Approach: Parameterized Density
and Temperature Profiles}

For a spherically symmetric galaxy where the hot ISM is supported by
ideal gas pressure, equation (\ref{eqn.he}) can be solved for the
total gravitating mass (stars, gas, and DM) as a function of the gas
density $(\rho_{\rm gas})$ and temperature  $(T)$,
\begin{equation}
M(<r) = -\left[\frac{rk_{\rm B}T}{\mu m_{\rm a}G}\right] \left[
\frac{d\ln\rho_{\rm gas}}{d\ln r} + \frac{d\ln T}{d\ln r} \right], \label{eqn.trad}
\end{equation}
where $k_{\rm B}$ is Boltzmann's constant, $\mu$ is the mean atomic
weight of the gas, $m_{\rm a}$ is the atomic mass unit, and $G$ is
Newton's gravitation constant. Beginning with M~87 over thirty years
ago~\citep{math78a,fabr80a}, this equation has been widely applied to
galaxies and clusters traditionally by inserting smooth parameterized
functions for the density and temperature into equation
(\ref{eqn.trad}).  The ad hoc parameterized functions do not guarantee
a physical mass profile that increases monotonically with increasing
radius, and their use offsets any gain in rigor from not assuming a
specific mass model.  Thus, while this traditional approach has the
advantage of being conceptually simple and computationally
inexpensive, it is best employed as a rough estimate and a check on
other methods.

\paragraph{\bf Smoothing by Binning}

The derivatives in equation (\ref{eqn.trad}) may instead be computed
directly from the density and temperature values measured in adjacent
radial bins rather than by assuming any parameterized models. Because
of bin-to-bin statistical fluctuations in the density and temperature
(or pressure) profiles, increased binning (i.e., smoothing) of the
radial profile is generally required to produce an approximate
monotonically increasing mass profile.  This method has only been
attempted for the brightest galaxy clusters~\citep{davi01a,voig06a},
since only for data of the highest quality can interesting results be
obtained without resorting to heavy binning of the data.

Recently, Nulsen et al.~\citep{nuls10a} have proposed a scheme that
obviates the need for explicit computation of the derivatives by
assuming that within any spherical shell the gravitating matter
density and gas temperature are constant.  By requiring that the gas
pressure be continuous across shell boundaries, the gas density is
then computed within any shell using an isothermal solution of
equation (\ref{eqn.he}). But the pressure continuity condition
involves the difference in the gravitational potential between
adjacent shells, which is related to a derivative (and is the
argument of an exponential). Consequently, the mass profile derived
from this procedure remains subject to similar bin-to-bin statistical
fluctuations that occur when the derivatives in equation
(\ref{eqn.trad}) are computed explicitly. To improve the smoothness of
the derived mass profile, Nulsen et al.\ advocate requiring that the
gravitating matter density decrease monotonically with increasing
radius.

\paragraph{\bf Bayesian Inversion with Smoothing Prior}

A very general approach to smoothed inversion has been proposed by Das
et al.\ \citep{das10a}, which is based on the seminal works of Merritt
\& Tremblay~\citep{merr94a} and Magorrian~\citep{mago99a}. Das et al.\
enforce smoothness via a Bayesian prior which penalizes large second
derivatives between radial bins in the temperature, circular velocity
(hence the mass), and pressure gradient. The magnitude of the
smoothing is controlled by a parameter $\lambda$, the value of which is
arbitrary and depends on both the number of radial bins and the model
grid spacing.  For convenience Das et al.\ use simulations to select a
fixed $\lambda$ for all galaxies. Since the sensitivity of the derived
mass profile to $\lambda$ should be investigated for any galaxy, a
reasonable extension would be to define a prior for $\lambda$ and then
marginalize over $\lambda$ for each system.

\subsubsection{Forward-Fitting}

\paragraph{\bf Pros:}
\begin{itemize}
\item No formal restrictions on galaxy geometry (though spherical symmetry
usually assumed).  
\item Allows for self-consistent extrapolation of hydrostatic model
and projection of galaxy emission to and from any radii outside the data
range.
\end{itemize}

\paragraph{\bf Cons:}
\begin{itemize}
\item Requires an input parameterized model for the mass.
\item Requires an input parameterized model for one of the following
thermodynamic quantities:  gas temperature, density, pressure, or
entropy.
\end{itemize}

The forward-fitting approach uses parameterized models for the mass
and one thermodynamic variable of the hot ISM (density, temperature,
pressure, or entropy). Because the galaxy is completely specified by
the models, in principle any geometrical configuration can be
accommodated, though spherical symmetry is usually assumed.  The model
can be extrapolated to radii outside the available data range allowing
for self-consistent treatment of the projection of the ISM of the
entire galaxy.  Forward-fitting can be applied to data that have been
deprojected (as with smoothed inversion) or directly to the data on
the sky. The ability to fit projected models to data on the sky is
especially important for lower quality, noisy data whose integrity
would be unacceptably degraded by deprojection noise.

This approach is preferred when it is desired to test the viability of
a particular mass model and to measure its parameters. If the assumed
model is incorrect, then the measured mass at a given radius may
differ systematically from the true value. It is therefore necessary
to explore multiple mass models to gauge the magnitude of any bias.
Since the measured mass may be sensitive to the assumed mass model,
when possible results using forward-fitting should be compared to
those from smoothed inversion.

Prior to \chandra\ and \xmm, when accurate spatially resolved
temperature profiles were only available for a few of the most massive
galaxies, it was customary to make strong assumptions about the
temperature profile in order to obtain an interesting constraint on
the DM. Typically the gas was assumed to be isothermal or that it
obeyed a polytropic equation of state $(P_{\rm gas} \propto \rho_{\rm
gas}^{\gamma})$. The most widely used of these approaches is the
well-known isothermal ``beta model'' \citep{cava76a,sara77a,beta}.
While such procedures still can be useful for obtaining DM estimates
from observations of low data quality, we focus our discussion on
methods intended to exploit accurate spatially resolved temperature
profiles, as such data are now more widely available.

\paragraph{\bf Density-Based}

If an assumption is made about the form of the gas density profile,
e.g., that it is described by a parameterized function, then taken
together with an assumed parameterized mass model the equation of
hydrostatic equilibrium (equation \ref{eqn.he}) may be solved to give
the temperature,
\begin{equation}
T(r)= T_0\frac{\rho_{\rm gas, 0}}{\rho_{\rm gas}(r)} - \frac{\mu
m_{\rm a} G}{k_{\rm B}}\frac{1}{\rho_{\rm gas}(r)} \int_{r_0}^{r}
\rho_{\rm gas}(r)\frac{M(<r)}{r^2} \, dr,
\label{eqn.density}
\end{equation}
where $T_0 = T(r_0)$ and $\rho_{\rm gas, 0}=\rho_{\rm gas}(r_0)$
evaluated at some reference radius $r_0$. Here and below we assume
spherical symmetry unless noted otherwise (i.e., in \S
\ref{nonsph}). The parameters of the mass and density models and the
normalization $T_0$ are then constrained by the observation. As with
all the types of forward-fitting approaches we discuss, it is common
practice first to measure the temperature and density profiles from
fitting the spectra and then constrain the free parameters of the
hydrostatic model via a simultaneous fit to the measured profiles.
Although the gas contribution to the total mass $M(<r)$ in elliptical
galaxies can generally be neglected until one approaches close to the
virial radius, a self-consistent hydrostatic model should include
it. For these density-based models this is straightforward since the
gas density profile is assumed, and therefore the gas mass profile is
defined before the temperature profile is computed.

An important variation of the density-based approach was proposed by
Fabian and colleagues~\citep{deproj,kris83,daw97a,alle01d} which
exploits the relatively high statistical quality of the X-ray surface
brightness profile.  Sometimes referred to as the ``Cambridge
Method'', this approach assumes spherical symmetry and analytically
deprojects the surface brightness using the onion peel procedure to
give the three-dimensional emissivity profile, $\epsilon \propto
\rho_{\rm gas}^2\Lambda(T,Z)$, where $\Lambda(T,Z)$ is the plasma
emissivity which depends on temperature $T$ and metal abundances $Z$
and is integrated over the relevant (broad) wave band.  If
$\Lambda(T,Z)$ is assumed to be constant over the galaxy, then
$\rho_{\rm gas}(r)\propto \sqrt{\epsilon(r)}$. Because of the
relatively high statistical quality of the surface brightness (and
emissivity), the density profile derived in this way usually can be
evaluated in much finer spatial bins than is possible for the
temperature profile determined from detailed spectral fitting.  This
allows the finely binned density profile to be treated as a continuous
function by interpolating values between the bins, which can then be
inserted (along with a parameterized mass model) into equation
(\ref{eqn.density}) to predict the temperature profile for comparison
to the values measured from the spectra.

It is a significant advantage of the Cambridge Method that it does not
require for input an assumed, possibly oversimplified, parameterized
model for the density profile.  Still, the density profile generated
assuming a constant $\Lambda$ leads to hydrostatic solutions that are
not self-consistent in the presence of temperature and metallicity
gradients. This effect is not expected to be large because variations
in $\Lambda$ within a galaxy or cluster are probably no more than
$\sim 20\%$ which translate to inferred density differences of only
$\sim 10\%$.  In principle, the best-fitting temperature profile could
be used to define a new spatially varying $\Lambda$ from which a new
density profile could be computed. The process could be iterated to
achieve improved self-consistency.

\paragraph{\bf Temperature-Based}

If an assumption is made about the form of the temperature profile,
then in conjunction with a parameterized mass profile the equation of
hydrostatic equilibrium (equation \ref{eqn.he}) may be solved for
the gas density,
\begin{equation}
\rho_{\rm gas}(r) = \rho_{\rm gas, 0}\frac{T_0}{T(r)}\exp\left[
-\frac{\mu m_{\rm a} G}{k_{\rm B}}\int_{r_0}^r\frac{1}{T(r)}\frac{M(<r)}{r^2}dr\right],
\label{eqn.temp}
\end{equation}
where spherical symmetry is again assumed. This approach can be
especially useful when the temperature profile is poorly constrained
by the data. Simple profiles (constant, power-law) can be adopted in
order to explore their impact on the inferred mass profile.  Unlike
the density-based case, the gas mass contribution to $M(<r)$ cannot be
included fully self-consistently in the fit.  However, because the gas
mass can be treated as a small perturbation, an iterative procedure
can be employed; i.e., after first taking the gas mass contribution to
$M(<r)$ to be zero, use the resultant best-fitting $\rho_{\rm gas}(r)$
to compute a fixed gas mass profile that is added to $M(<r)$ and then
fit again.

\paragraph{\bf Entropy-Based}

Recent studies have suggested that the entropy is the logical
thermodynamic variable for which to assume a parameterized
model~\citep{hump08a,hump09c,cava09a,fusc09a}.  Consistent with the
hydrostatic equilibrium approximation is the requirement that the hot
gas be stable against convection at every radius. Convective stability
provides an important additional constraint on the mass profile that
has been used in the past to set a robust lower limit on the total
masses of galaxy clusters~\citep{fabi86a}. For an ideal gas equation
of state the first law of thermodynamics can be solved to yield the
specific entropy, $s = (3 k_{\rm B}/2\mu m_{\rm a})\ln(T\rho_{\rm
gas}^{-2/3})\rm + constant.$ It is conventional for studies of the hot
gas in galaxies and clusters to remove the logarithm and the numerical
factor to define the quantity, $S \equiv (k_{\rm B}/\mu m_{\rm
a})T\rho_{\rm gas}^{-2/3},$ which we will refer to as the ``entropy''
in our discussion.

If parameterized models are assumed for both the entropy and mass
profiles, then the equation of hydrostatic equilibrium can be
rewritten so that,
\begin{equation}
\frac{d\xi}{dr} = -\frac{2}{5}\frac{GM(<r)}{r^2}S^{-3/5}, \hskip 1cm
\xi \equiv P^{2/5}. \label{eqn.entropy}
\end{equation}
Solving this equation for $\xi$ gives the profiles of gas density,
$\rho_{\rm gas}= (P/S)^{3/5} = S^{-3/5}\xi^{3/2}$, and temperature,
$k_{\rm B}T/\mu m_{\rm a} = S^{3/5}P^{2/5} = S^{3/5}\xi$, which are
compared to the observation to constrain the parameters of the input
$S$ and $M(<r)$ models.  Since, however, the total mass $M(<r)$ also
includes the gas mass which depends on $\xi$, equation
(\ref{eqn.entropy}) can be solved for $\xi$ by direct integration only
if the gas mass is negligible with respect to the mass of the stars
and DM.  To include the gas mass self-consistently, equation
(\ref{eqn.entropy}) can be differentiated and rearranged to yield,
\begin{equation}
\frac{d}{dr}\left(r^2S^{3/5}\frac{d\xi}{dr}\right) + \frac{8\pi
r^2G}{5}S^{-3/5}\xi^{3/2} = -\frac{8\pi r^2G}{5}
\left(\rho_{\rm stars} + \rho_{\rm DM}\right),
\end{equation}
where use has been made of the relation $dM(<r)/dr = 4\pi r^2
(\rho_{\rm stars} + \rho_{\rm DM} + \rho_{\rm gas}).$ The two
boundary conditions for this second-order differential equation
require specifying the values of $\xi$ and $d\xi/dr$ at some
radius. The value of $\xi$ amounts to specifying the pressure at some
radius, which is just a normalization constant determined by the
fitting of the density and temperature profiles.  As discussed
by~\citep{hump08a}, because the gas mass makes a negligible contribution
to the total mass at small radius, equation (\ref{eqn.entropy}) can be
evaluated at such a small radius to give the $d\xi/dr$ boundary condition.

The Schwarzschild criterion for convective stability requires that the
entropy increases monotonically with increasing radius, consistent
with observations of relaxed elliptical galaxies, for which the
entropy profiles are well described by simple models of a constant
with a broken power-law~\citep{hump08a,hump09c,hump11a}.  The
Schwarzschild criterion also effectively limits the magnitude of any
temperature gradient.  Stronger restrictions on the temperature
gradient potentially could be enforced by MTI/HBI
instabilities~\citep{balb01a,quat08a} that develop in the weakly
magnetized plasma characteristic of the hot gas in a galaxy
cluster. These instabilities arise when heat and momentum transport
occurs only along magnetic field lines, which may happen in regions of
a plasma where the gyroradius is much smaller than the the Coulomb
mean free path (such as for an elliptical galaxy -- see \S
\ref{pre}).  But since these instabilities are expected to be
weaker for the lower temperatures of elliptical galaxies, and they are
also suppressed by small amounts of
turbulence~\citep{parr10a,rusz10a}, their relevance to elliptical
galaxies is unclear.

Whereas gas density and temperature profiles display a large range of
behavior, a robust result of cosmological simulations considering only
gravity is that the radial entropy profile follows a power-law,
$S(r)\sim r^{1.1}$~\citep{tozz01a,voit05a}. Since the cooling and
feedback processes that would modify this relation are expected to be
most significant closer to the centers of halos, the $S(r)\sim
r^{1.1}$ relation provides a physically well-motivated asymptotic
relation that is particularly useful when extrapolating the
hydrostatic model outside the data range (e.g., computing model
projections).

\section{Is DM Required?}

\begin{figure*}[t]
\centerline{\includegraphics[scale=0.80,angle=-90]{NGC1332_temperature.ps}}

\vskip 0.3cm

\centerline{\includegraphics[scale=0.80,angle=-90]{NGC4261_temperature.ps}}
\caption{\label{fig.tprof} 
\chandra\ temperature profiles of two galaxies
from~\citep{hump09c}. The top
panel shows the centrally rising profile of NGC~1332 characteristic of
galaxies with $M_{\rm vir}$~\ltsim~$10^{13}\, M_{\odot}$ and average
temperature $k_{\rm B} T\sim 0.5$~keV.  The bottom panel shows the
centrally falling profile of NGC~4261 characteristic of galaxies with
$M_{\rm vir}$~\gtsim~$10^{13}\, M_{\odot}$ and average temperature
$k_{\rm B} T\sim 1$~keV. The $K$-band half-light radii are 2.7 and
3.4~kpc, respectively, for NGC~1332 and NGC~4261.}
\end{figure*}

\begin{figure*}[t]

\parbox{0.49\textwidth}{
\centerline{\includegraphics[scale=0.29,angle=0]{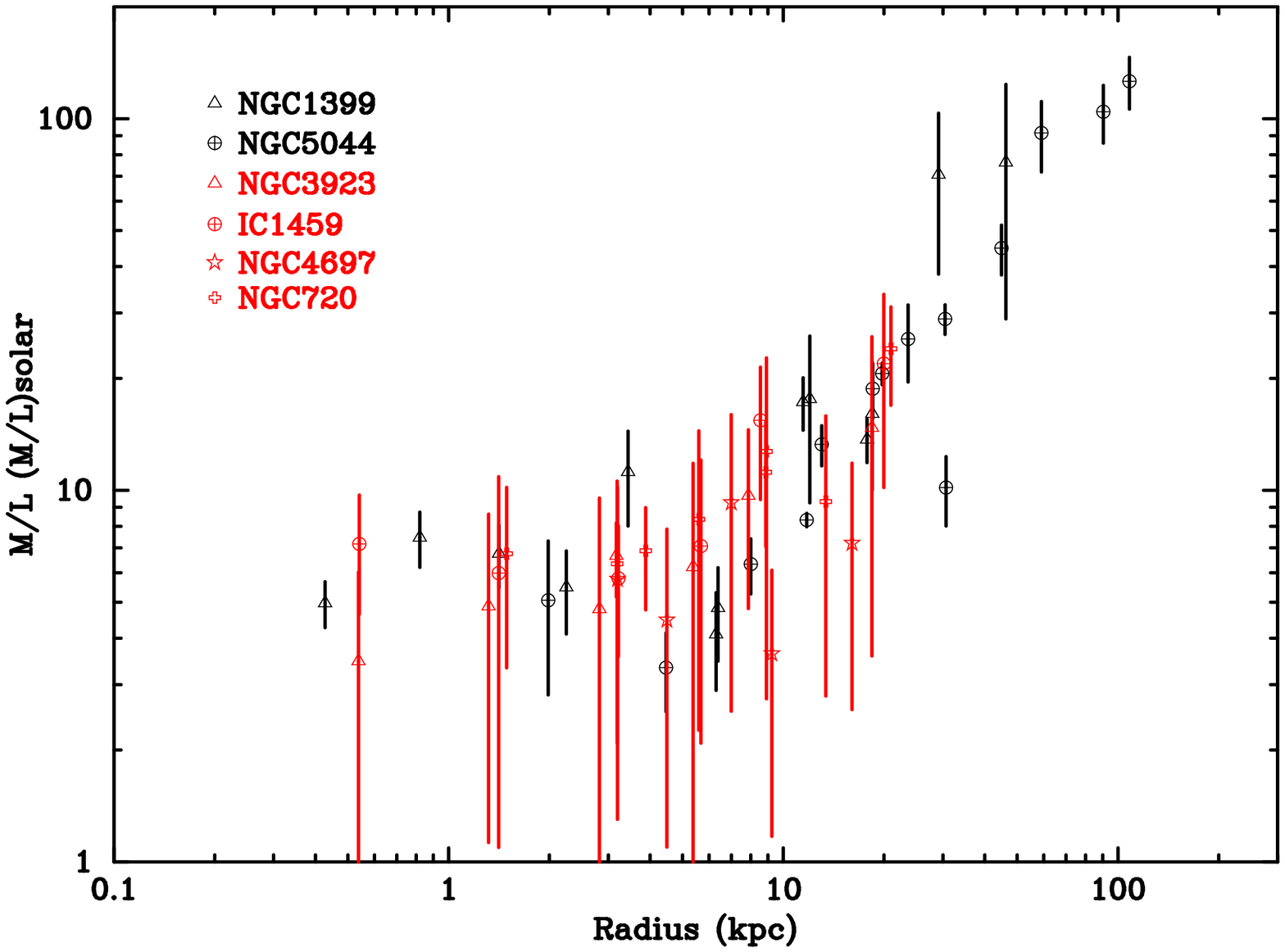}}}
\parbox{0.49\textwidth}{ \vspace{1.25cm}
\centerline{\includegraphics[scale=0.25,angle=-90]{h06_mass_to_light_all.ps}}}

\vskip 0.3cm

\parbox{0.49\textwidth}{
\centerline{\includegraphics[scale=0.28,angle=0]{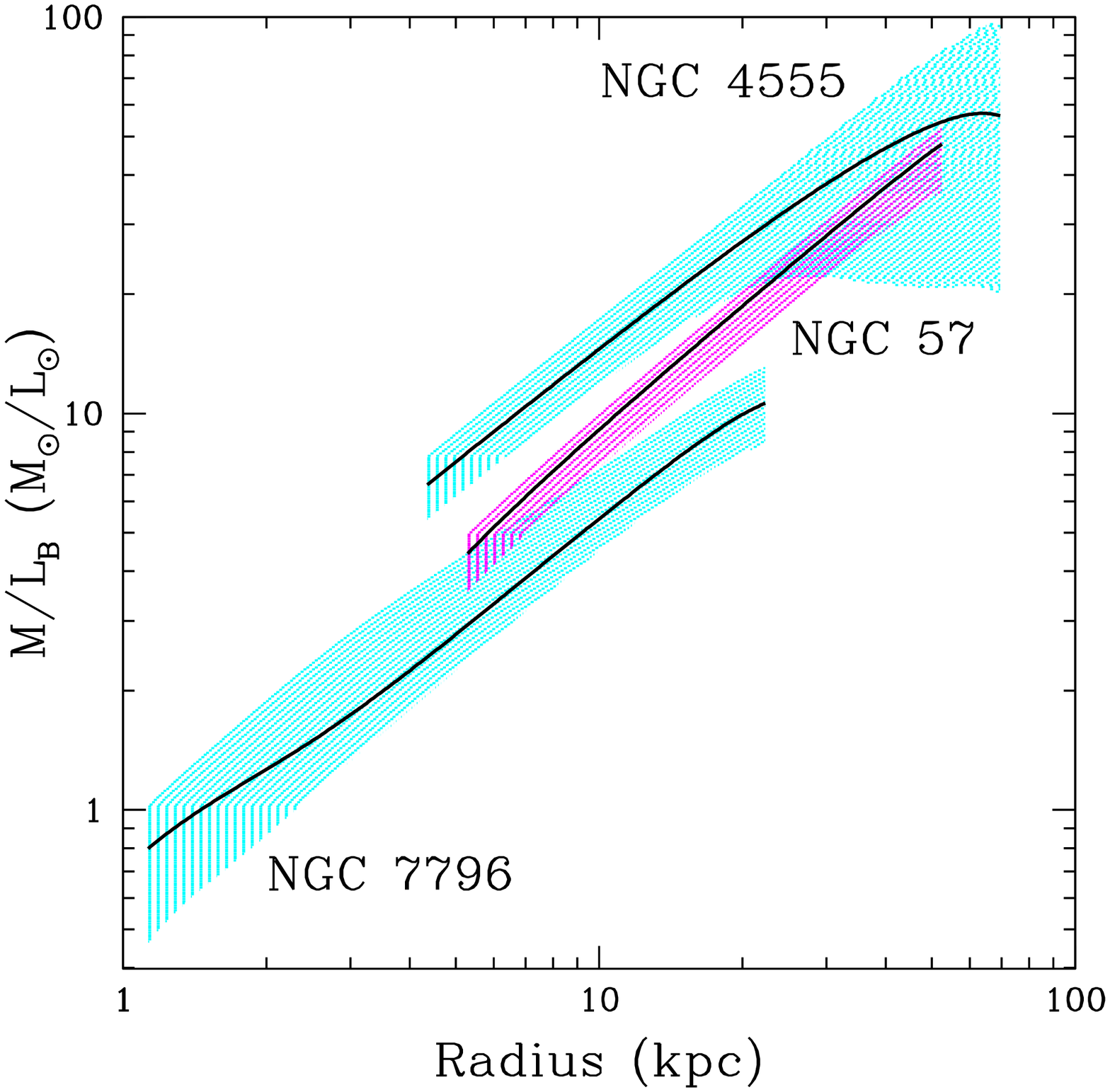}}}
\parbox{0.49\textwidth}{
\centerline{\includegraphics[scale=0.25,angle=0]{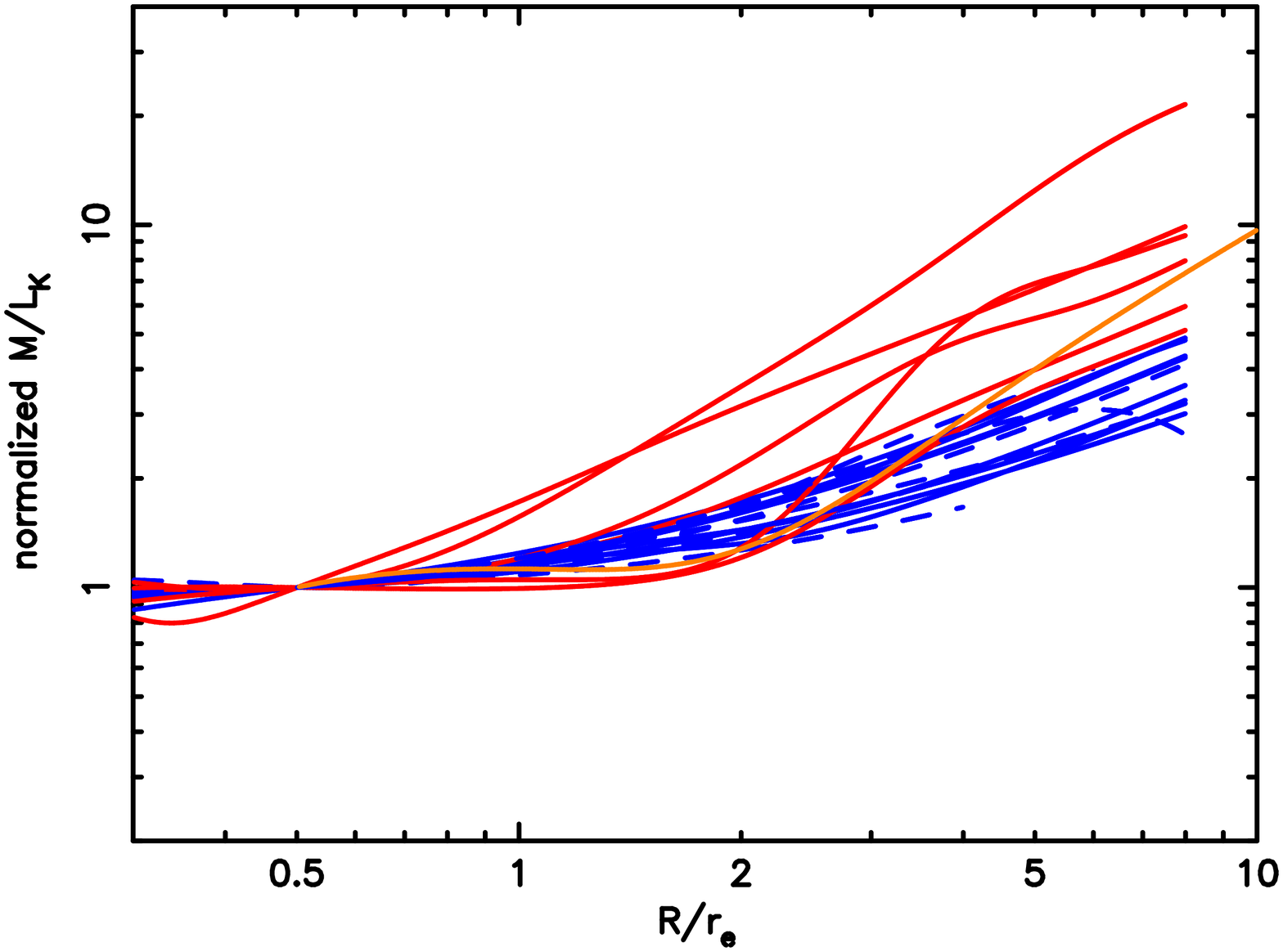}}}
\caption{\label{fig.ml}
Representative total mass-to-light ($M/L$) ratios of elliptical galaxies obtained from
\chandra\ and \xmm\ observations clearly showing rising $M/L$, and
thus the need for DM, outside of $1-2$ optical half-light radii.
({\em Top Left}) Clear confirmation of rising $M/L$ in the massive
galaxies NGC~1399 and NGC~5044~\citep{fuka06a} found previously with
\rosat\ observations.  ({\em Top Right}) Results for the galaxies
displayed in Figure~\ref{fig.h06} obtained by~\citep{hump06b}. The
black regions correspond to the galaxies NGC~720, NGC~4325, and
NGC~6482 that have $M_{\rm vir}$~\ltsim~$10^{13}\, M_{\odot}$ while
the red regions correspond to the galaxies NGC~1407, NGC~4472,
NGC~4649, and NGC~4261 that have $M_{\rm vir}$~\gtsim~$10^{13}\,
M_{\odot}$. Each result shown reflects a $1\sigma$ confidence region.
({\em Bottom Left}) $1\sigma$ error regions for three galaxies
from~\citep{osul04a,osul07a}, where isothermal temperature profiles
have been adopted for NGC~57 and NGC~7796. ({\em Bottom Right})
Results from~\citep{nagi09a} for $M/L$ profiles of galaxies with
temperature profiles that fall toward the center (red lines) and for
others (blue lines).}
\end{figure*}

\begin{figure*}[t]
\centerline{\includegraphics[scale=0.39,angle=-90]{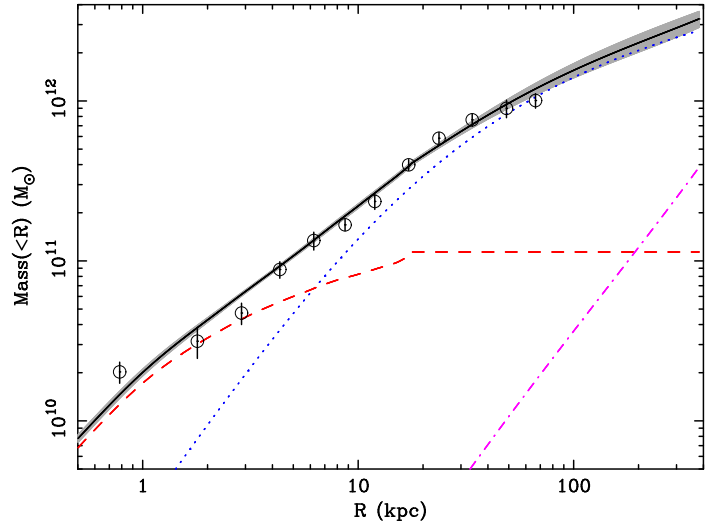}}
\caption{\label{fig.n720.mass}
Radial mass profile of NGC~720 obtained by~\citep{hump11a} using deep
observations with \chandra\ and \suzaku. The solid (black) line
indicates the total enclosed mass, the dashed (red) line indicates the
stellar mass, the dotted (blue) line is the DM contribution, and the
dash-dot (magenta) line is the gas mass contribution. The grey shaded
regions indicate the $1\sigma$ error on the total mass distribution
obtained using the entropy-based forward-fitting method. Overlaid are
a set of data points derived from an independent analysis using the
traditional smoothed inversion approach; i.e., the model indicated by
the shaded region was not obtained by fitting these data
points. Notice that DM is clearly required for radii larger than $\sim
5$~kpc, about $2R_e$, where $R_e=3.1$~kpc is the $K$-band half-light
radius.}
\end{figure*}

Prior to the launches of \chandra\ and \xmm, the principal limiting
factor for most X-ray studies of DM in elliptical galaxies was the
temperature profile. (Non-spherical DM constraints, which are largely
insensitive to the temperature profile, can be more affected by
unresolved discrete sources~\citep{buot96d,buot97a,buot02b} -- see \S
\ref{nonsph}). In most cases isothermality was assumed, but the
possibility of large radial temperature gradients meant that no firm
conclusions on DM could be obtained for most
galaxies~\citep{fabb89}. For a small number of the brightest, hottest
($k_{\rm B} T\sim 1$~keV) most massive ($\rm few
\times 10^{13}\msun$) galaxies, \rosat\ was able to place interesting
constraints on the temperature profiles, from which large amounts of
DM were inferred: NGC~1339~\citep{jone97a},
NGC~4472~\citep{irwi96a,brig97a}, NGC~4649~\citep{brig97a}, and
NGC~5044~\citep{davi94}.

Now accurate temperature profiles have been measured with \chandra\
and \xmm\ for many galaxies (Figure \ref{fig.tprof}). The shapes of
the radial temperature profiles of early-type galaxies can be
separated into roughly two classes~\citep{fuka06a,hump06b,nagi09a}:
(1) those with profiles that rise toward the center, corresponding to
galaxies with $M_{\rm vir}$~\ltsim~$10^{13}\, M_{\odot}$ and average
temperature $k_{\rm B} T\sim 0.5$~keV, and (2) those with profiles
that fall toward the center, corresponding to galaxies with $M_{\rm
vir}$~\gtsim~$10^{13}\, M_{\odot}$ and average temperature $k_{\rm B}
T\sim 1$~keV. These temperature profiles enable accurate DM
constraints for many galaxies, including lower mass systems ($\rm few
\times 10^{12}\msun$).

Equipped with accurate temperature profiles from \chandra\ and \xmm,
precise constraints on mass-to-light ratios ($M/L$) have now been
obtained for many
galaxies~\citep{sun03a,osul04a,fuka06a,hump06b,gast07b,osul07a,osul07b,zhan07a,hump08a,hump09c,nagi09a,hump11a};
here $M$ is the total mass and $L$ is the stellar luminosity, each
computed within the same radius.  Generally, for those galaxies with
good constraints on the temperature profile out to a large radius, the
data exclude a constant $M/L$ profile at high significance, instead
requiring that $M/L$ rises with increasing radius outside of 1-2
optical half-light radii (Figure \ref{fig.ml}). The failure of the
constant $M/L$ model provides strong evidence for DM.

It should be stressed that the current generation of X-ray satellites is
able for the first time to provide excellent constraints on DM in
lower mass galaxies with average temperature $k_{\rm B} T\sim
0.5$~keV. Combining data from deep observations with \chandra\ and
\suzaku, \citep{hump11a} obtained strong constraints on the mass
profile of NGC~720 (see Figure \ref{fig.n720.mass}) which was found to
have a total mass $(M_{\rm 2500}=1.6\pm 0.2\times 10^{12}\,
M_{\odot})$ similar to that of the Milky Way.

This X-ray evidence for DM determined individually for a relatively
small number of elliptical galaxies ($\sim 15$, depending on the upper
mass limit) is complemented by evidence from large statistical studies
in the optical. Weak lensing~\citep{hoek05a,klein06a,mand06a,heym06a}
and the dynamics of satellite galaxies~\citep{prad03a} consider the
accumulated data from a very large number of galaxies and find that
the average halo mass $\sim 10^{12}\, M_{\odot}$ clearly exceeds that
which can be attributed to the stars. As for constraints on individual
systems, strong evidence for DM using sophisticated, orbit-based
stellar dynamical models has been found recently for
NGC~4649~\citep{shen10a}, although the mass somewhat exceeds that
inferred by X-rays, a point to which we shall return in more detail in
\S \ref{acc}.

\section{Radial DM Profile}
\label{rad}

Although the need for DM is now established for many elliptical
galaxies, the X-ray data do not as of yet clearly favor a specific
model for the radial profile. In particular, both the NFW and $\sim
r^{-2}$ DM density profiles are consistent with available data.  The
NFW profile provides a useful means to compare observed DM halos to
those predicted by cosmological models. The two parameters which
specify the NFW profile are the concentration $c_{\Delta}$ and halo
mass $M_{\Delta}$.  The concentration is defined as,
$c_{\Delta}=r_{\Delta}/r_s$, where $r_s$ is a scale radius (denoting
the radius at which the logarithmic denstiy slope is -2), and
$r_{\Delta}$ is a reference radius defined so that the average density
of the halo within the sphere of radius $r_{\Delta}$ equals the number
$\Delta$ times the critical density of the Universe. Quoted values for
$\Delta$ typically range from 100-2500. The mass, $M_{\Delta}$, is the
mass enclosed within $r_{\Delta}$.  Often $\Delta$ is chosen to be the
solution to the spherical collapse model at the time of cluster
virialization~\citep{brya98a}, where $\Delta\sim 100$ at the present
epoch in the standard $\Lambda$CDM cosmology, in which case the
parameters are sometimes referred to as the ``virial radius'', $r_{\rm
vir}$, ``virial mass'', $M_{\rm vir}$, and ``virial concentration'',
$c_{\rm vir}$.

\begin{figure*}[t]
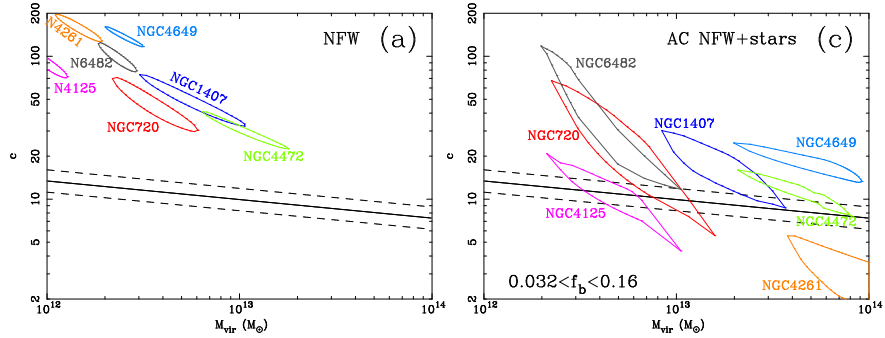

\parbox{0.49\textwidth}{
\centerline{\includegraphics[scale=0.24,angle=-90]{fill_dm_f4a.eps}}}
\parbox{0.49\textwidth}{ 
\centerline{\includegraphics[scale=0.24,angle=-90]{fill_dm_f4c.eps}}}
\caption{\label{fig.cm} 
$c_{\rm vir}-M_{\rm vir}$ results ($1\sigma$) for the seven elliptical
galaxies shown in Figure \ref{fig.h06} from \citep{hump06b} where the
mass model includes ({\em Left Panel}) only a single NFW component and
({\em Right Panel}) an NFW model for the DM and a Hernquist model for
for the stellar mass. The omission of the stellar component leads to a
large overestimate of the concentration parameter as originally
suggested by~\citep{mamo05a}. When including components for both the
stars and DM the values are mostly consistent with the theoretical
expectation within the errors (solid line is median and dashed lines
$1\sigma$ intrinsic scatter), with a weak suggestion that the galaxies
lie systematically above the theoretical relation.}
\end{figure*}

Dissipationless simulations of the $\Lambda$CDM model predict $c_{\rm
vir}\approx 10$ for DM halos with an intrinsic scatter of $\approx
0.1$~dex over the mass range of elliptical galaxies considered in this
review~\citep{bull01a,macc08a}.  Some early measurements obtained much
larger concentrations on these mass scales inconsistent with
theory. Firstly, \citep{sato00a} attempted to deconvolve the large,
energy-dependent PSF of \asca\ observations, which is challenging
given that the size of the PSF ($>1^{\prime}$) is generally as large
or larger than the optical half-light radii of nearby elliptical
galaxies observed in X-rays. Secondly, \citep{wu00a} used \rosat\
observations, but they had to assume the gas is isothermal throughout
each galaxy. Finally, one early study using \chandra\ data of
NGC~6482~\citep{khos04a} with an accurately measured (non-isothermal)
temperature profile also obtained a very large concentration
($c_{200}=61, c_{\rm vir}\sim 80$).
 
Mamon \& Lokas~\citep{mamo05a} proposed that such large concentrations
disagreeing with theoretical expectation were primarily an artifact of
neglecting to include a component for the stellar mass in X-ray
determinations of the DM profile. This bias was confirmed with
\chandra\ observations by \citep{hump06b} in their sample of seven
galaxies (Figure \ref{fig.cm}), demonstrating that reliable X-ray
measurements of the DM concentration on the galaxy scale definitely
require the stellar mass to be modeled accurately.  There is good
agreement in the concentrations obtained for studies of galaxies like
NGC~1407 that include the stellar mass~\citep{hump06b,zhan07a}.  But
statistically significant systematic differences can arise in some
(higher mass) galaxies between studies that include stellar
mass~\citep{gast07b} and those that instead exclude a large portion of
the central region~\citep{sun09a}. As emphasized by~\citep{gast07b},
since an accurate determination of the DM concentration requires that
the NFW scale radius be accurately measured, it is essential that
$r_s$ lie within the range of data being fitted. One the other hand,
excluding the central region of a galaxy disturbed by AGN feedback is
a reasonable approach for hydrostatic analysis. In either case, the
systematic error in the adopted analysis choice needs to be
investigated and quantified.

The observed $c_{\rm vir}-M_{\rm vir}$ relation for elliptical
galaxies inferred from X-rays (Figure \ref{fig.cm}) is broadly
consistent with the cosmological prediction within the estimated
observational errors. Despite the relatively large error regions,
these measurements on the elliptical galaxy scale, when combined with
measurements for groups and massive clusters, were crucial to
providing a sufficiently wide mass baseline to establish that the
$c_{\rm vir}-M_{\rm vir}$ slope is indeed negative (a robust
prediction of CDM models) with a value $(-0.17\pm 0.03$) close to the
theoretical prediction~\citep{buot07a}.  However, even when accounting
for the stellar mass the normalization of the $c_{\rm vir}-M_{\rm
vir}$ relation on the galaxy scale remains larger than
predicted. While the effect is not highly significant for the results
shown in Figure~\ref{fig.cm}, the recent analysis of
NGC~720~\citep{hump11a} using deeper \chandra\ data mentioned
previously with an improved entropy-based analysis finds $c_{\rm vir}
= 26\pm 5$, $M_{\rm vir} = 3.3\pm 0.4\,
\times 10^{12}\, M_{\odot}$, which exceeds the theoretical prediction
(considering intrinsic scatter) by about $2\sigma$.  The systematic
offset could represent one or more of the following: the selection of
preferentially early forming systems (NGC~720 is a very relaxed
isolated galaxy), departures from hydrostatic equilibrium
(underestimate of stellar mass contribution), or the need for a
different value of $\sigma_8$ or $w$ in the cosmological
model~\citep[see][]{buot07a}. Although adiabatic
contraction~\citep{blum84a,gned04a} can lower the inferred
concentrations, the estimated modest reductions still exceed the
theoretical prediction~\citep{hump06b,gast07b,hump11a}.

\begin{figure*}[t]
\centerline{\includegraphics[scale=0.59,angle=0]{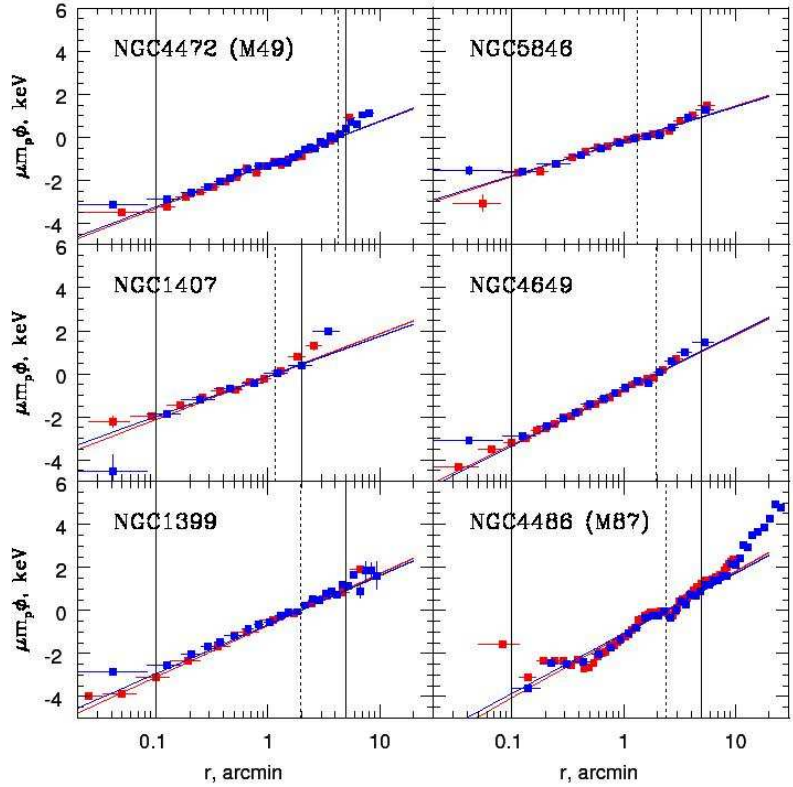}}
\caption{\label{fig.pot}
Gravitational potential profiles and power-law fits
from~\citep{chur10a}. \chandra\ data are shown in red and \xmm\ in
blue. Solid vertical lines indicate the radial range included for the
power-law fits, while the vertical dotted lines indicate the optical
half-light radius.}
\end{figure*}

\begin{figure*}[t]
\parbox{0.49\textwidth}{
\centerline{\includegraphics[scale=0.23,angle=-90]{h10a_f1.eps}}}
\parbox{0.49\textwidth}{ 
\centerline{\includegraphics[scale=0.31,angle=0]{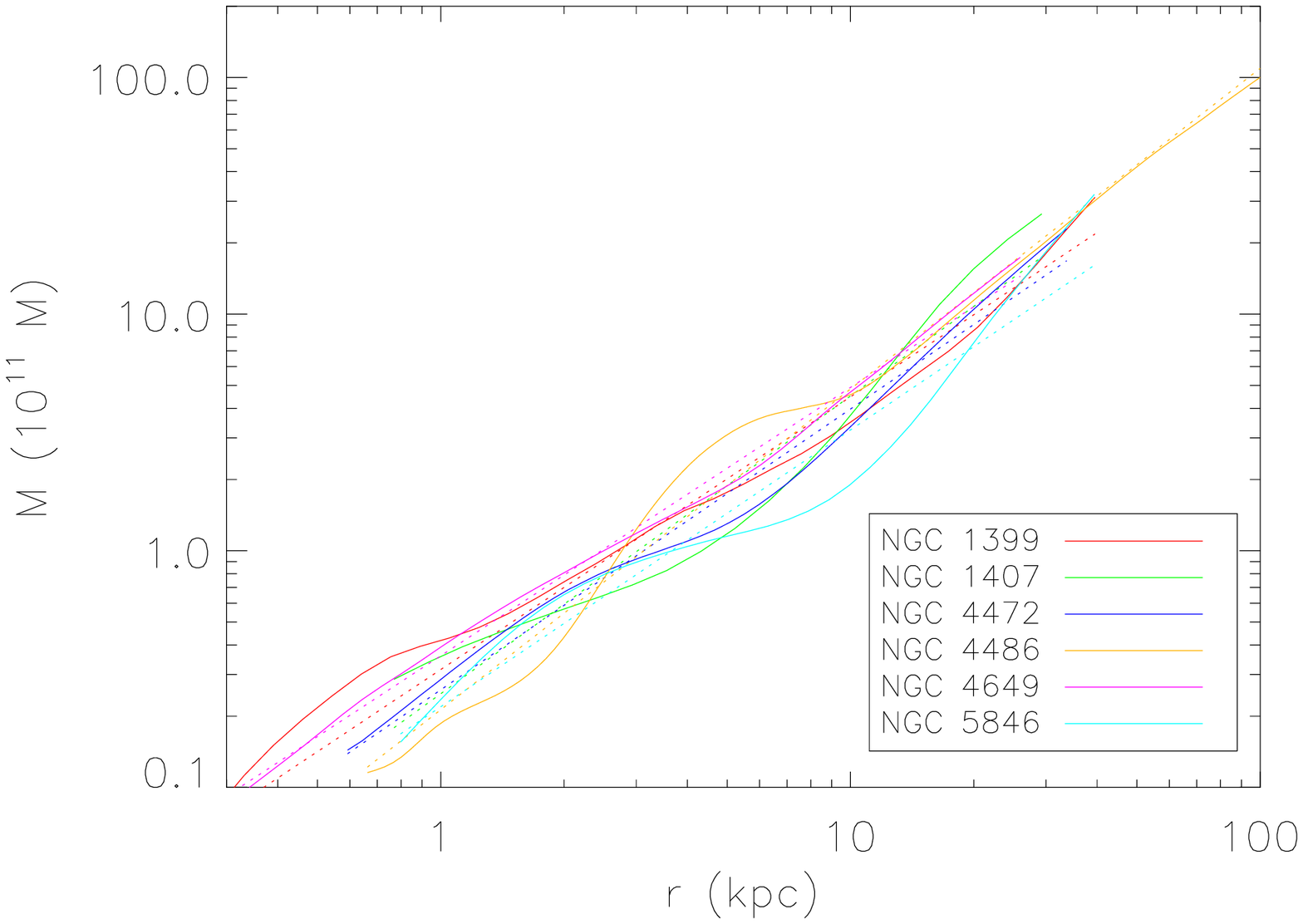}}}
\caption{\label{fig.powlaw} 
Radial mass profiles and associated power-law fits. ({\em Left Panel})
Results for 10 objects spanning galaxies to clusters
from~\citep{hump10a}, arbitrarily scaled for clarity. The solid lines
are the best-fitting profiles determined from an entropy-based
forward-fitting procedure, while the data points are determined from
the traditional smoothed inversion approach; i.e., the displayed
models are not fitted to the data points but are derived
independently. ({\em Right Panel}) Results for 6 objects
from~\citep{das10a} based on X-ray data analyzed
by~\citep{chur10a}. Solid lines show the mass best-fitting mass
profiles, while the dotted lines show the best-fitting power-laws to
these mass profiles.}
\end{figure*}

\begin{figure*}[t]
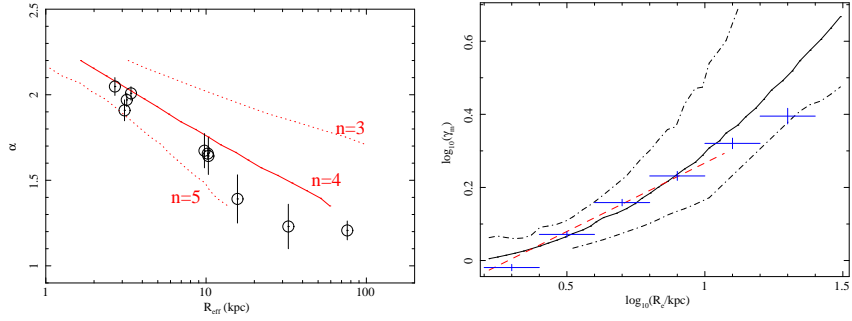

\parbox{0.49\textwidth}{
\centerline{\includegraphics[scale=0.23,angle=-90]{h10a_f2a_nolensing.eps}}}
\parbox{0.49\textwidth}{ 
\centerline{\includegraphics[scale=0.23,angle=-90]{h10a_f5.eps}}}
\caption{\label{fig.h10a} 
From~\citep{hump10a}: ({\em Left Panel}) X-ray (black circles)
measurements of logarithmic mass slope plotted against optical
half-light radius. The red lines are model predictions decomposing a
power-law total mass profile into Sersic stellar light and NFW DM
profiles (see text).  ({\em Right Panel}) Ratio $\gamma_m$ of total
mass to stellar mass enclosed within $R_e$ plotted vs.\ $R_e$. The
solid line is the model prediction corresponding to the $n=4$ curve
shown in the left panel which agrees well with the data points
corresponding to the $K$-band $M/L$ ratios adapted from the data of
\citep{laba08a}.}
\end{figure*}

While the NFW profile is a good description of the DM profiles in
elliptical galaxies, evidence continues to accumulate that the total
gravitating mass is very well approximated by a single power-law over
a wide radial range. Prior to \chandra\ and \xmm\ many X-ray studies
found mass profiles consistent with $\rho\sim r^{-\alpha}$, where
$\alpha\approx 2$ (see~\citep{chur10a,hump10a}, and references
therein), although isothermal gas was generally assumed. These results
have now been confirmed and strengthened (Figures~\ref{fig.pot},
\ref{fig.powlaw}) with accurate temperature profiles measured with
\chandra\ and \xmm\
\citep{fuka06a,hump06b,chur10a,hump10a,das10a}.  Furthermore, using \chandra\
data of 10 galaxies spanning a wide range in $\mvir$ (including some
clusters) and optical half-light radii ($R_e$), \citep{hump10a}
discovered that the power-law index $\alpha$ decreases systematically
with increasing $R_e$ (Figure \ref{fig.h10a}). This behavior can be
explained by the combination of the stellar (Sersic) and DM (NFW)
profiles required to produce a power-law total mass profile, which
leads to the relation $\alpha = 2.31 - 0.54\log_{10} R_e$ for a Sersic
index $n\approx 4$ corresponding to a stellar de Vaucouleurs profile.

This correlation has recently been confirmed by~\citep{auge10a} using
combined central velocity dispersions and strong gravitational lensing
in 73 early-type galaxies. These optical results provide very
important supporting evidence, but since they essentially determine a
mass profile from two mass data points (i.e., at the center and at the
Einstein radius of a galaxy), the X-ray measurements remain of
critical importance.  This relation between $\alpha$ with $R_e$ has
far-reaching implications, as \citep{hump10a} showed that it implies a
DM fraction within $R_e$ that varies systematically with the
properties of the galaxy in such a manner as to reproduce, without
fine tuning, the observed tilt of the Fundamental Plane. Consequently,
\citep{hump10a} speculated that establishing a nearly power-law total mass
distribution is a fundamental feature of galaxy formation and the
primary factor which determines the tilt of the fundamental plane.

\section{Baryon Fraction}

Please see the full article to be published in the book, {\sl Hot
Interstellar Matter in Elliptical Galaxies}, eds.\ D.-W.\ Kim \& S.\
Pellegrini, Astrophysics \& Space Science Library (ASSL), Springer.

\section{Non-Spherical Constraints}
\label{nonsph}

Please see the full article to be published in the book, {\sl Hot
Interstellar Matter in Elliptical Galaxies}, eds.\ D.-W.\ Kim \& S.\
Pellegrini, Astrophysics \& Space Science Library (ASSL), Springer.

\section{SMBHs}
\label{smbh}

Please see the full article to be published in the book, {\sl Hot
Interstellar Matter in Elliptical Galaxies}, eds.\ D.-W.\ Kim \& S.\
Pellegrini, Astrophysics \& Space Science Library (ASSL), Springer.

\section{Accuracy of the Hydrostatic Equilibrium Approximation}
\label{acc}

\begin{figure*}[t]
\centerline{\includegraphics[scale=0.33,angle=270]{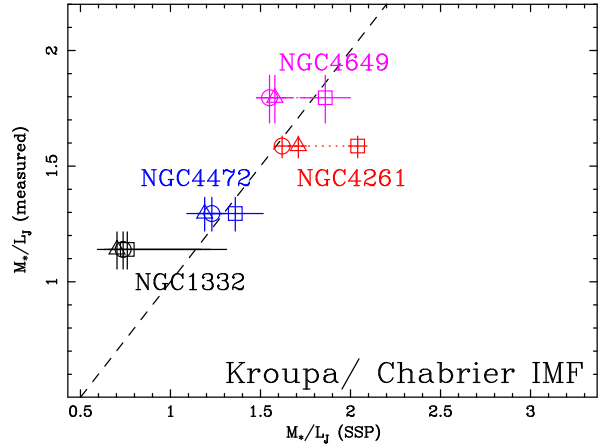}}
\caption{ \label{fig.mlstars} 
Comparison of the stellar M/L ratios from X-ray mass modeling with
those predicted from three sets of SSP models (Squares:
\citep{mara05}; circles: \citep{fioc97}; triangles:
\citep{bruz03}). The dotted line is ``y=x''. The good agreement
implies deviations from hydrostatic equilibrium are less than
$\sim$10--20\%. Adapted from \citep{hump09c}.}
\end{figure*}

Provided other sources of systematic error are controlled, the
accuracy of a mass measurement from X-rays reflects the degree to
which non-thermal motions are present in the hot gas. As remarked in
\S \ref{pre} only limited direct constraints on gas motions are possible
with current X-ray detectors, and such measurements that exist suggest
subsonic non-thermal motions.  An indirect assessment of non-thermal
motions can be achieved by comparing the X-ray--determined stellar
$M/L$ ratios with those predicted for each galaxy by single-burst
stellar population synthesis (SSP) models. For a sample of 4 galaxies
(including NGC~4472 and NGC~4261, both of which have AGN-blown
cavities in the ISM), the X-rays and SSP display good overall
agreement (Fig~\ref{fig.mlstars}), implying non-thermal pressure is no
larger than $\sim$20\% \citep{hump09c}. Similar agreement was found
for NGC~1407~\citep{zhan07a}. While the SSP models could be in error
due to a mixture of differently aged stars in each galaxy or the
initial mass function (IMF) differing from that of
Kroupa~\citep{krou01a}, for non-thermal pressure to be significant, it
must exist in a finely balanced conspiracy with the shape of the IMF.

Alternatively, if the three-dimensional gravitational potential can be
measured independently, it should be possible to use the measured
X-ray properties of the gas to make an (indirect) constraint on
non-thermal gas motions.  Since elliptical galaxies with X-ray data of
sufficient quality for detailed mass analysis are necessarily quite
nearby (within $\sim$ 100~Mpc), the only techniques (aside from X-ray
tools) currently available to measure the gravitational potential with
sufficient accuracy are axisymmetric (or triaxial) stellar dynamical
methods.  The most sophisticated of these methods employ the full
information in the line-of-sight stellar velocity distribution to
break the mass-orbit
degeneracy~\citep{vand93a,gebh00b,vand08a,gebh09a}, while making
minimal assumptions about the orbital
structure~\citep{schw79a,vand98a}.  The fundamentally ill-posed nature
of the inversion problem at the heart of these ``orbit-based'' methods
can be mitigated by regularization or a maximum entropy
constraint~\citep[\eg][]{vand98a,rich88a,gebh00b}.

\begin{figure*}
\parbox{0.49\textwidth}{
\centerline{\includegraphics[scale=0.31,angle=0]{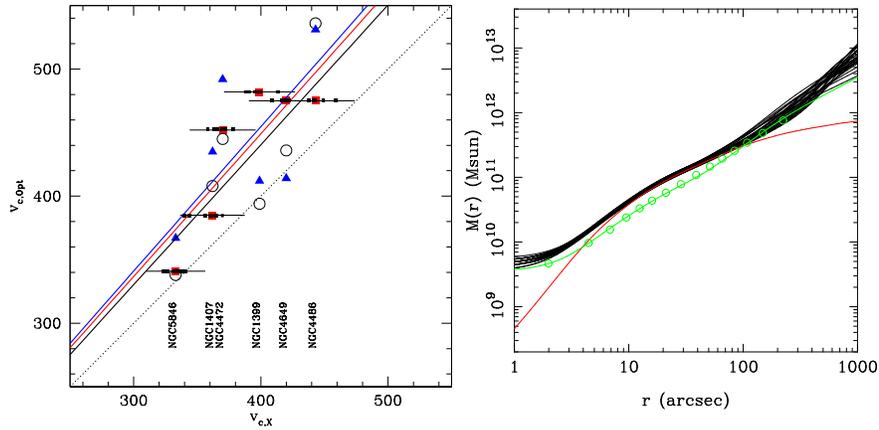}}}
\parbox{0.49\textwidth}{
\centerline{\includegraphics[scale=0.32,angle=-90]{shen_gebhardt_f4.eps}}}
\caption{\footnotesize \label{fig.sd} 
({\sl Left Panel}) Comparison of circular velocities for six nearby
galaxies inferred from hydrostatic X-ray models (horizontal axis) and
from a suite of simple stellar dynamical estimates (vertical
axis). The observation that the galaxies tend to lie above the dotted
line representing equal X-ray and optical values implies that the
X-ray method tends to underestimate the mass by around 30\%, implying
significant non-thermal gas motions~\citep{chur10a}. ({\sl Right
Panel}) Stellar dynamics/X-ray comparison for NGC~4649
from~\citep{shen10a}. The black lines represent the total mass profile
(and $1\sigma$ range) from the stellar dynamics. The green line is a
representation of the mass profile inferred from
X-rays~\citep{hump08a}, and the red line is the stellar mass.}
\end{figure*}

Only a few studies have attempted to combine the X-ray and optical
methods in this way.  Assuming spherical symmetry and severely
limiting the orbital structure by employing ``Jeans modeling'', a
modest disagreement between the mass inferred from the two techniques
has been reported for the galaxy NGC~4472~\citep{math03c,ciot04a}.
While this may indicate on average $\sim$20--50\%\ non-thermal
pressure support, it could also reflect the limitations of Jeans
modeling to recover the orbital structure
accurately~\citep[e.g.,][]{binn90a}.  In contrast, using a similar
approach~\citep{roma09a} argued the X-ray method overestimates the
mass distribution in the elliptical galaxy NGC~1407 by $\sim$70\%,
which is difficult to understand unless the gas is globally
out-flowing. The exact magnitude of this discrepancy, however, is very
uncertain due to large systematic errors associated with assumptions
in both the X-ray and the optical analysis. (A similar conclusion
obtained for the galaxy NGC~3379~\citep{pell06a} is also very
uncertain since unresolved point sources dominate the X-ray emission
in that particular galaxy~\citep{revn08a,trin08a}.)  Large systematic
uncertainties have also characterized similar studies of 6 galaxies
undertaken by~\citep{chur08a} and~\citep{chur10a}, who concluded that,
on average, the X-ray data are consistent with modest, but highly
uncertain, non-thermal support (see Figure~\ref{fig.sd}). They
conclude that, although they find $\sim$30\%\ non-thermal pressure in
the systems they studied, ``{\em the uncertainties in [their] model
assumptions (e.g., spherical symmetry) are sufficiently large that the
contribution [of non-thermal pressure] could be consistent with
zero.}'' \citep{chur08a}. By an extension of this argument, it could
also be significantly larger.

A major roadblock to the use of more rigorous axisymmetric (or
triaxial) ``orbit based'' methods for modeling the optical data in
such work has been the prohibitive computational cost of
self-consistently incorporating a DM halo in such a scheme. Recently,
however, such models have begun to be
developed~\citep[e.g.,][]{thom07a}. In particular,
\citep{shen10a} found a discrepancy between the mass profile of
NGC\thin 4649 from orbit-based models and the (spherical) X-ray mass
distribution, implying on average $\sim$40\%\ non-thermal
support. This is significantly larger than was required by the
hydrodynamical models of \citep{brig09a} to reconcile the stellar and
X-ray ellipticity profiles (although this comparison is not strictly
self-consistent as the hydrodynamical models assumed a gravitational
potential consistent with the X-ray results). Similarly,
\citep{gebh09a} found discrepancies for the galaxy M\thin 87, while
the X-ray mass profiles were found by \citep{das10a} to underestimate
by $\sim$20--45\%\ the gravitating mass determined from orbit-based
stellar models in the centers of three galaxies, with the sign of the
discrepancy reversing at large (\gtsim 10~kpc) radii.  While
inclination effects could conceivably contribute a significant
uncertainty into the stellar dynamical analysis of the round galaxies
considered in these studies, a similar mismatch between the X-ray and
(orbit-based) optical mass estimates (corresponding to $\sim$50\%\
non-thermal support) has been reported in the central $\sim$3~kpc
region for the edge-on S0 galaxy NGC~1332 \citep{rusl10a}.

Hence, the current status of comparisons between stellar dynamics and
X-rays is that X-rays tend, on average, to underestimate the mass
obtained by stellar dynamics by $\sim 30\%$, but with considerable
uncertainty on many of the measurements, and the comparison can be
complicated by a radial dependence in the discrepancy.  To better
understand the systematic errors in these comparisons, it is essential
in the future that self-consistent comparisons be made; i.e.,
simultaneous modeling of the X-ray and stellar dynamical data.  The
X-ray data analysis is subject to numerous potential systematic
effects which need to be controlled; e.g., background
characterization, measurements of the metal abundances, treatment of
unresolved discrete sources, and even uncertainties associated
with the mass modeling technique that is
adopted~\citep[e.g.,][]{hump09c,das10a}.  While many of these
systematic errors are modest in absolute terms, they can be comparable
to, or at times exceed, the statistical errors. With the detailed
comparison between X-ray and optical mass profiles now demanding the
interpretation of differences as small as $\sim$20\%, precise control
of these errors is essential. Until this becomes routine, the accurate
radial profiles of possible non-thermal gas motions inferred by
comparing mass profiles determined from X-rays and stellar dynamics
will remain out of reach.


\section{Conclusions}

With the advent of accurate radial temperature profiles from \chandra,
\xmm, and \suzaku, the need for DM is now firmly established in
many giant elliptical galaxies $(\rm 10^{12}\, M_{\odot} \lta
M_{vir}\lta 10^{13}\, M_{\odot})$. Current data do not distinguish
between NFW and isothermal DM profiles.  For the cosmologically
motivated NFW model the inferred concentration parameters of
elliptical galaxies generally exceed the mean theoretical
relation. This discrepancy may be due in large part to the
preferential selection of the most relaxed galaxies for X-ray studies.
X-ray observations confirm that the total mass profile (baryons+DM) is
close to isothermal $(M\sim r)$, and new evidence suggests a more
general power-law relation for the slope of the total mass profile
that varies with the stellar half-light radius. While the global
baryon fraction remains difficult to constrain, a recent joint
\chandra-\suzaku\ analysis of NGC~720 indicates a global baryon
fraction consistent with the cosmological value, suggesting a Milky
Way--size galaxy can retain all of its baryons. The axial ratio of the
DM inferred from the X-ray isophote shapes of NGC~720 is very
consistent with those of halos produced in cosmological
simulations. Finally, the unprecedented high spatial resolution of
\chandra\ has enabled the first measurements of the mass of SMBHs in a
few elliptical galaxies, obtaining a precision comparable to that
achieved by traditional optical methods in NGC~4649.

The most important uncertainty in X-ray determinations of DM in
elliptical galaxies is no longer the measurement of the temperature
profile but now the accuracy of the hydrostatic equilibrium
approximation; i.e., the magnitude of non-thermal motions in the hot
ISM. The limited direct constraints possible with current data suggest
subsonic non-thermal motions, and improved measurements of turbulent
motions via Doppler line broadening in the brightest systems are
expected from {\sl Astro-H} in a few years.  In the meantime, it is
essential to perform self-consistent studies of X-ray and stellar
dynamical data of elliptical galaxies to assess the magnitude of
non-thermal gas motions and achieve more reliable DM measurements than
possible with either technique individually.

\begin{acknowledgement}
We thank W.\ Mathews, F.\ Brighenti, S.\ Ettori, and K.\ Gebhardt for
discussions related to this work. We are grateful to W.\ Mathews and
F.\ Gastaldello for providing comments on the manuscript. We also
would like to express our appreciation to E.\ O'Sullivan for providing
data included in Figure \ref{fig.ml}. This work was supported in part
by the National Aeronautics and Space Administration under Grant No.\
NNX10AD07G issued through the Astrophysics Data Analysis Program
(ADP).
\end{acknowledgement}
%


\begin{thebibliography}{100}
\providecommand{\url}[1]{{#1}}
\providecommand{\urlprefix}{URL }
\expandafter\ifx\csname urlstyle\endcsname\relax
  \providecommand{\doi}[1]{DOI \discretionary{}{}{}#1}\else
  \providecommand{\doi}{DOI \discretionary{}{}{}\begingroup
  \urlstyle{rm}\Url}\fi

\bibitem{feng05a}
J.L. {Feng}, ArXiv High Energy Physics - Phenomenology e-prints  (2005)

\bibitem{hann06a}
S.~{Hannestad}, International Journal of Modern Physics A \textbf{21}, 1938
  (2006).
\newblock \doi{10.1142/S0217751X06032885}

\bibitem{eina09a}
J.~{Einasto}, ArXiv e-prints  (2009)

\bibitem{stef09a}
F.D. {Steffen}, European Physical Journal C \textbf{59}, 557 (2009).
\newblock \doi{10.1140/epjc/s10052-008-0830-0}

\bibitem{iras09a}
I.G. {Irastorza}, ArXiv e-prints  (2009)

\bibitem{ellij10a}
J.~{Ellis}, K.A. {Olive}, ArXiv e-prints  (2010)

\bibitem{pdg08}
C.~Amsler, et~al., Phys. Lett. \textbf{B667}, 1 (2008).
\newblock \doi{10.1016/j.physletb.2008.07.018}

\bibitem{laha10a}
O.~{Lahav}, A.R. {Liddle}, ArXiv e-prints  (2010)

\bibitem{moor94a}
B.~{Moore}, \nat \textbf{370}, 629 (1994).
\newblock \doi{10.1038/370629a0}

\bibitem{flor94a}
R.A. {Flores}, J.R. {Primack}, \apjl \textbf{427}, L1 (1994).
\newblock \doi{10.1086/187350}

\bibitem{gilm08a}
G.~{Gilmore}, D.~{Zucker}, M.~{Wilkinson}, R.F.G. {Wyse}, V.~{Belokurov},
  J.~{Kleyna}, A.~{Koch}, N.W. {Evans}, E.K. {Grebel}, in \emph{Astronomical
  Society of the Pacific Conference Series}, \emph{Astronomical Society of the
  Pacific Conference Series}, vol. 399, ed. by {T.~Kodama, T.~Yamada, \&
  K.~Aoki} (2008), \emph{Astronomical Society of the Pacific Conference
  Series}, vol. 399, p. 453

\bibitem{prim09a}
J.R. {Primack}, New Journal of Physics \textbf{11}(10), 105029 (2009).
\newblock \doi{10.1088/1367-2630/11/10/105029}

\bibitem{debl10a}
W.J.G. {de Blok}, Advances in Astronomy \textbf{2010} (2010).
\newblock \doi{10.1155/2010/789293}

\bibitem{sand02b}
D.J. {Sand}, T.~{Treu}, R.S. {Ellis}, \apjl \textbf{574}, L129 (2002).
\newblock \doi{10.1086/342530}

\bibitem{kels02a}
D.D. {Kelson}, A.I. {Zabludoff}, K.A. {Williams}, S.C. {Trager}, J.S.
  {Mulchaey}, M.~{Bolte}, \apj \textbf{576}, 720 (2002).
\newblock \doi{10.1086/341891}

\bibitem{sand04b}
D.J. {Sand}, T.~{Treu}, G.P. {Smith}, R.S. {Ellis}, \apj \textbf{604}, 88
  (2004).
\newblock \doi{10.1086/382146}

\bibitem{sand08a}
D.J. {Sand}, T.~{Treu}, R.S. {Ellis}, G.P. {Smith}, J.~{Kneib}, \apj
  \textbf{674}, 711 (2008).
\newblock \doi{10.1086/524652}

\bibitem{newm09a}
A.B. {Newman}, T.~{Treu}, R.S. {Ellis}, D.J. {Sand}, J.~{Richard}, P.J.
  {Marshall}, P.~{Capak}, S.~{Miyazaki}, \apj \textbf{706}, 1078 (2009).
\newblock \doi{10.1088/0004-637X/706/2/1078}

\bibitem{arab02a}
J.S. {Arabadjis}, M.W. {Bautz}, G.P. {Garmire}, \apj \textbf{572}, 66 (2002).
\newblock \doi{10.1086/340296}

\bibitem{lewi03a}
A.D. {Lewis}, D.A. {Buote}, J.T. {Stocke}, \apj \textbf{586}, 135 (2003)

\bibitem{sper00}
D.N. {Spergel}, P.J. {Steinhardt}, Physical Review Letters \textbf{84}, 3760
  (2000).
\newblock \doi{10.1103/PhysRevLett.84.3760}

\bibitem{koch00a}
C.S. {Kochanek}, M.~{White}, \apj \textbf{543}, 514 (2000).
\newblock \doi{10.1086/317149}

\bibitem{firm01a}
C.~{Firmani}, E.~{D'Onghia}, G.~{Chincarini}, X.~{Hern{\'a}ndez},
  V.~{Avila-Reese}, \mnras \textbf{321}, 713 (2001).
\newblock \doi{10.1046/j.1365-8711.2001.04030.x}

\bibitem{ahn05a}
K.~{Ahn}, P.R. {Shapiro}, \mnras \textbf{363}, 1092 (2005).
\newblock \doi{10.1111/j.1365-2966.2005.09492.x}

\bibitem{rand08a}
S.W. {Randall}, M.~{Markevitch}, D.~{Clowe}, A.H. {Gonzalez}, M.~{Brada{\v c}},
  \apj \textbf{679}, 1173 (2008).
\newblock \doi{10.1086/587859}

\bibitem{kuzi10a}
R.~{Kuzio de Naray}, G.D. {Martinez}, J.S. {Bullock}, M.~{Kaplinghat}, \apjl
  \textbf{710}, L161 (2010).
\newblock \doi{10.1088/2041-8205/710/2/L161}

\bibitem{blum84a}
G.R. {Blumenthal}, S.M. {Faber}, J.R. {Primack}, M.J. {Rees}, \nat
  \textbf{311}, 517 (1984)

\bibitem{elza04a}
A.A. {El-Zant}, Y.~{Hoffman}, J.~{Primack}, F.~{Combes}, I.~{Shlosman}, \apjl
  \textbf{607}, L75 (2004).
\newblock \doi{10.1086/421938}

\bibitem{gned07a}
O.Y. {Gnedin}, D.H. {Weinberg}, J.~{Pizagno}, F.~{Prada}, H.~{Rix}, \apj
  \textbf{671}, 1115 (2007).
\newblock \doi{10.1086/523256}

\bibitem{roma08a}
E.~{Romano-D{\'{\i}}az}, I.~{Shlosman}, Y.~{Hoffman}, C.~{Heller}, \apjl
  \textbf{685}, L105 (2008).
\newblock \doi{10.1086/592687}

\bibitem{delp09a}
A.~{Del Popolo}, \apj \textbf{698}, 2093 (2009).
\newblock \doi{10.1088/0004-637X/698/2/2093}

\bibitem{duff10a}
A.R. {Duffy}, J.~{Schaye}, S.T. {Kay}, C.~{Dalla Vecchia}, R.A. {Battye}, C.M.
  {Booth}, \mnras \textbf{405}, 2161 (2010).
\newblock \doi{10.1111/j.1365-2966.2010.16613.x}

\bibitem{abad10a}
M.G. {Abadi}, J.F. {Navarro}, M.~{Fardal}, A.~{Babul}, M.~{Steinmetz}, \mnras
  p. 847 (2010).
\newblock \doi{10.1111/j.1365-2966.2010.16912.x}

\bibitem{ferr00a}
L.~{Ferrarese}, D.~{Merritt}, \apjl \textbf{539}, L9 (2000).
\newblock \doi{10.1086/312838}

\bibitem{gebh00a}
K.~{Gebhardt}, R.~{Bender}, G.~{Bower}, A.~{Dressler}, S.M. {Faber}, A.V.
  {Filippenko}, R.~{Green}, C.~{Grillmair}, L.C. {Ho}, J.~{Kormendy}, T.R.
  {Lauer}, J.~{Magorrian}, J.~{Pinkney}, D.~{Richstone}, S.~{Tremaine}, \apjl
  \textbf{539}, L13 (2000).
\newblock \doi{10.1086/312840}

\bibitem{birn03a}
Y.~{Birnboim}, A.~{Dekel}, \mnras \textbf{345}, 349 (2003).
\newblock \doi{10.1046/j.1365-8711.2003.06955.x}

\bibitem{kere05a}
D.~{Kere{\v s}}, N.~{Katz}, D.H. {Weinberg}, R.~{Dav{\'e}}, \mnras
  \textbf{363}, 2 (2005).
\newblock \doi{10.1111/j.1365-2966.2005.09451.x}

\bibitem{birn07a}
Y.~{Birnboim}, A.~{Dekel}, E.~{Neistein}, \mnras \textbf{380}, 339 (2007).
\newblock \doi{10.1111/j.1365-2966.2007.12074.x}

\bibitem{math03a}
W.G. {Mathews}, F.~{Brighenti}, \araa \textbf{41}, 191 (2003)

\bibitem{deke08a}
A.~{Dekel}, Y.~{Birnboim}, \mnras \textbf{383}, 119 (2008).
\newblock \doi{10.1111/j.1365-2966.2007.12569.x}

\bibitem{brig09a}
F.~{Brighenti}, W.G. {Mathews}, P.J. {Humphrey}, D.A. {Buote}, \apj
  \textbf{705}, 1672 (2009).
\newblock \doi{10.1088/0004-637X/705/2/1672}

\bibitem{ponm94}
T.J. {Ponman}, D.J. {Allan}, L.R. {Jones}, M.~{Merrifield}, I.M. {McHardy},
  H.J. {Lehto}, G.A. {Luppino}, \nat \textbf{369}, 462 (1994).
\newblock \doi{10.1038/369462a0}

\bibitem{vikh99a}
A.~{Vikhlinin}, B.R. {McNamara}, A.~{Hornstrup}, H.~{Quintana}, W.~{Forman},
  C.~{Jones}, M.~{Way}, \apjl \textbf{520}, L1 (1999).
\newblock \doi{10.1086/312134}

\bibitem{jone03a}
L.R. {Jones}, T.J. {Ponman}, A.~{Horton}, A.~{Babul}, H.~{Ebeling}, D.J.
  {Burke}, \mnras \textbf{343}, 627 (2003).
\newblock \doi{10.1046/j.1365-8711.2003.06702.x}

\bibitem{dong05a}
E.~{D'Onghia}, J.~{Sommer-Larsen}, A.D. {Romeo}, A.~{Burkert}, K.~{Pedersen},
  L.~{Portinari}, J.~{Rasmussen}, \apjl \textbf{630}, L109 (2005).
\newblock \doi{10.1086/491651}

\bibitem{dari07a}
A.~{Dariush}, H.G. {Khosroshahi}, T.J. {Ponman}, F.~{Pearce},
  S.~{Raychaudhury}, W.~{Hartley}, \mnras \textbf{382}, 433 (2007).
\newblock \doi{10.1111/j.1365-2966.2007.12385.x}

\bibitem{milo06a}
M.~{Milosavljevi{\'c}}, C.J. {Miller}, S.R. {Furlanetto}, A.~{Cooray}, \apjl
  \textbf{637}, L9 (2006).
\newblock \doi{10.1086/500547}

\bibitem{fabb89}
G.~{Fabbiano}, \araa \textbf{27}, 87 (1989).
\newblock \doi{10.1146/annurev.aa.27.090189.000511}

\bibitem{buot98a}
D.A. {Buote}, C.R. {Canizares}, in \emph{Galactic Halos}, \emph{Astronomical
  Society of the Pacific Conference Series}, vol. 136, ed. by {D.~Zaritsky}
  (1998), \emph{Astronomical Society of the Pacific Conference Series}, vol.
  136, p. 289

\bibitem{gerh06a}
O.~{Gerhard}, in \emph{Planetary Nebulae Beyond the Milky Way}, ed. by
  {L.~Stanghellini, J.~R.~Walsh, \& N.~G.~Douglas} (2006), p. 299.
\newblock \doi{10.1007/3-540-34270-2-47}

\bibitem{gerh10a}
O.~{Gerhard}, ArXiv e-prints  (2010)

\bibitem{elli10a}
R.S. {Ellis}, Phil.\ Trans.\ R.\ Soc.\ A \textbf{368}, 967 (2010).
\newblock \doi{10.1098/rsta.2009.0209}

\bibitem{treu10a}
T.~{Treu}, \araa \textbf{48}, 87 (2010).
\newblock \doi{10.1146/annurev-astro-081309-130924}

\bibitem{sara86a}
C.L. {Sarazin}, Reviews of Modern Physics \textbf{58}, 1 (1986).
\newblock \doi{10.1103/RevModPhys.58.1}

\bibitem{fabi90a}
A.C. {Fabian}, in \emph{NATO ASIC Proc. 300: Physical Processes in Hot Cosmic
  Plasmas}, ed. by {W.~Brinkmann, A.~C.~Fabian, \& F.~Giovannelli} (1990), pp.
  271--297

\bibitem{sara92a}
C.L. {Sarazin}, in \emph{NATO ASIC Proc. 366: Clusters and Superclusters of
  Galaxies}, ed. by {A.~C.~Fabian} (1992), p. 131

\bibitem{mewe99a}
R.~{Mewe}, in \emph{X-Ray Spectroscopy in Astrophysics}, \emph{Lecture Notes in
  Physics, Berlin Springer Verlag}, vol. 520, ed. by {J.~van Paradijs \&
  J.~A.~M.~Bleeker} (1999), \emph{Lecture Notes in Physics, Berlin Springer
  Verlag}, vol. 520, p. 109

\bibitem{spit62a}
L.~{Spitzer}, \emph{{Physics of Fully Ionized Gases}} (1962)

\bibitem{spit56a}
L.~{Spitzer}, Jr., \apj \textbf{124}, 20 (1956).
\newblock \doi{10.1086/146200}

\bibitem{cowi77a}
L.L. {Cowie}, C.F. {McKee}, \apj \textbf{211}, 135 (1977).
\newblock \doi{10.1086/154911}

\bibitem{dopi03a}
M.A. {Dopita}, R.S. {Sutherland}, \emph{{Astrophysics of the diffuse universe}}
  (2003)

\bibitem{vall04a}
J.P. {Vall{\'e}e}, New Astronomy Review \textbf{48}, 763 (2004).
\newblock \doi{10.1016/j.newar.2004.03.017}

\bibitem{math97a}
W.G. {Mathews}, F.~{Brighenti}, \apj \textbf{488}, 595 (1997).
\newblock \doi{10.1086/304728}

\bibitem{govo04b}
F.~{Govoni}, L.~{Feretti}, International Journal of Modern Physics D
  \textbf{13}, 1549 (2004).
\newblock \doi{10.1142/S0218271804005080}

\bibitem{gast07b}
F.~{Gastaldello}, D.A. {Buote}, P.J. {Humphrey}, L.~{Zappacosta}, J.S.
  {Bullock}, F.~{Brighenti}, W.G. {Mathews}, \apj \textbf{669}, 158 (2007).
\newblock \doi{10.1086/521519}

\bibitem{sun09a}
M.~{Sun}, G.M. {Voit}, M.~{Donahue}, C.~{Jones}, W.~{Forman}, A.~{Vikhlinin},
  \apj \textbf{693}, 1142 (2009).
\newblock \doi{10.1088/0004-637X/693/2/1142}

\bibitem{wern09a}
N.~{Werner}, I.~{Zhuravleva}, E.~{Churazov}, A.~{Simionescu}, S.W. {Allen},
  W.~{Forman}, C.~{Jones}, J.S. {Kaastra}, \mnras \textbf{398}, 23 (2009).
\newblock \doi{10.1111/j.1365-2966.2009.14860.x}

\bibitem{sand10a}
J.S. {Sanders}, A.C. {Fabian}, R.K. {Smith}, ArXiv e-prints  (2010)

\bibitem{tsai94a}
J.C. {Tsai}, N.~{Katz}, E.~{Bertschinger}, \apj \textbf{423}, 553 (1994).
\newblock \doi{10.1086/173834}

\bibitem{evra96a}
A.E. {Evrard}, C.A. {Metzler}, J.F. {Navarro}, \apj \textbf{469}, 494 (1996).
\newblock \doi{10.1086/177798}

\bibitem{naga07a}
D.~{Nagai}, A.~{Vikhlinin}, A.V. {Kravtsov}, \apj \textbf{655}, 98 (2007).
\newblock \doi{10.1086/509868}

\bibitem{piff08a}
R.~{Piffaretti}, R.~{Valdarnini}, \aap \textbf{491}, 71 (2008).
\newblock \doi{10.1051/0004-6361:200809739}

\bibitem{fang09a}
T.~{Fang}, P.~{Humphrey}, D.~{Buote}, \apj \textbf{691}, 1648 (2009).
\newblock \doi{10.1088/0004-637X/691/2/1648}

\bibitem{caon00a}
N.~{Caon}, D.~{Macchetto}, M.~{Pastoriza}, \apjs \textbf{127}, 39 (2000).
\newblock \doi{10.1086/313315}

\bibitem{hanl00a}
P.C. {Hanlan}, J.N. {Bregman}, \apj \textbf{530}, 213 (2000).
\newblock \doi{10.1086/308357}

\bibitem{crai10a}
R.A. {Crain}, I.G. {McCarthy}, C.S. {Frenk}, T.~{Theuns}, J.~{Schaye}, \mnras
  p. 967 (2010).
\newblock \doi{10.1111/j.1365-2966.2010.16985.x}

\bibitem{crai10b}
R.A. {Crain}, I.G. {McCarthy}, J.~{Schaye}, C.S. {Frenk}, T.~{Theuns}, ArXiv
  e-prints  (2010)

\bibitem{buot94}
D.A. {Buote}, C.R. {Canizares}, \apj \textbf{427}, 86 (1994).
\newblock \doi{10.1086/174123}

\bibitem{buot96a}
D.A. {Buote}, C.R. {Canizares}, \apj \textbf{457}, 177 (1996).
\newblock \doi{10.1086/176721}

\bibitem{buot95a}
D.A. {Buote}, J.C. {Tsai}, \apj \textbf{439}, 29 (1995).
\newblock \doi{10.1086/175148}

\bibitem{hump08b}
P.J. {Humphrey}, D.A. {Buote}, \apj \textbf{689}, 983 (2008).
\newblock \doi{10.1086/592590}

\bibitem{fuka06a}
Y.~{Fukazawa}, J.G. {Botoya-Nonesa}, J.~{Pu}, A.~{Ohto}, N.~{Kawano}, \apj
  \textbf{636}, 698 (2006).
\newblock \doi{10.1086/498081}

\bibitem{pell06a}
S.~{Pellegrini}, L.~{Ciotti}, \mnras \textbf{370}, 1797 (2006).
\newblock \doi{10.1111/j.1365-2966.2006.10590.x}

\bibitem{revn08a}
M.~{Revnivtsev}, E.~{Churazov}, S.~{Sazonov}, W.~{Forman}, C.~{Jones}, \aap
  \textbf{490}, 37 (2008).
\newblock \doi{10.1051/0004-6361:200809889}

\bibitem{trin08a}
G.~{Trinchieri}, S.~{Pellegrini}, G.~{Fabbiano}, R.~{Fu}, N.J. {Brassington},
  A.~{Zezas}, D.~{Kim}, J.~{Gallagher}, L.~{Angelini}, R.L. {Davies},
  V.~{Kalogera}, A.R. {King}, S.~{Zepf}, \apj \textbf{688}, 1000 (2008).
\newblock \doi{10.1086/592287}

\bibitem{form07a}
W.~{Forman}, C.~{Jones}, E.~{Churazov}, M.~{Markevitch}, P.~{Nulsen},
  A.~{Vikhlinin}, M.~{Begelman}, H.~{B{\"o}hringer}, J.~{Eilek}, S.~{Heinz},
  R.~{Kraft}, F.~{Owen}, M.~{Pahre}, \apj \textbf{665}, 1057 (2007).
\newblock \doi{10.1086/519480}

\bibitem{mill10a}
E.T. {Million}, N.~{Werner}, A.~{Simionescu}, S.W. {Allen}, P.E.J. {Nulsen},
  A.C. {Fabian}, H.~{Bohringer}, J.S. {Sanders}, ArXiv e-prints  (2010)

\bibitem{chur08a}
E.~{Churazov}, W.~{Forman}, A.~{Vikhlinin}, S.~{Tremaine}, O.~{Gerhard},
  C.~{Jones}, \mnras \textbf{388}, 1062 (2008).
\newblock \doi{10.1111/j.1365-2966.2008.13507.x}

\bibitem{cava10a}
K.W. {Cavagnolo}, B.R. {McNamara}, P.E.J. {Nulsen}, C.L. {Carilli}, C.~{Jones},
  L.~{Birzan}, ArXiv e-prints  (2010)

\bibitem{hump08a}
P.J. {Humphrey}, D.A. {Buote}, F.~{Brighenti}, K.~{Gebhardt}, W.G. {Mathews},
  \apj \textbf{683}, 161 (2008).
\newblock \doi{10.1086/589709}

\bibitem{hump09c}
P.J. {Humphrey}, D.A. {Buote}, F.~{Brighenti}, K.~{Gebhardt}, W.G. {Mathews},
  \apj \textbf{703}, 1257 (2009).
\newblock \doi{10.1088/0004-637X/703/2/1257}

\bibitem{hump11a}
P.J. {Humphrey}, D.A. {Buote}, C.R. {Canizares}, A.C. {Fabian}, J.M. {Miller},
  \apj \textbf{729}, 53 (2011).
\newblock \doi{10.1088/0004-637X/729/1/53}

\bibitem{prat10a}
G.W. {Pratt}, M.~{Arnaud}, R.~{Piffaretti}, H.~{B{\"o}hringer}, T.J. {Ponman},
  J.H. {Croston}, G.M. {Voit}, S.~{Borgani}, R.G. {Bower}, \aap \textbf{511},
  A85 (2010).
\newblock \doi{10.1051/0004-6361/200913309}

\bibitem{hump06b}
P.J. {Humphrey}, D.A. {Buote}, F.~{Gastaldello}, L.~{Zappacosta}, J.S.
  {Bullock}, F.~{Brighenti}, W.G. {Mathews}, \apj \textbf{646}, 899 (2006).
\newblock \doi{10.1086/505019}

\bibitem{dieh07a}
S.~{Diehl}, T.S. {Statler}, \apj \textbf{668}, 150 (2007).
\newblock \doi{10.1086/521009}

\bibitem{jone02a}
C.~{Jones}, W.~{Forman}, A.~{Vikhlinin}, M.~{Markevitch}, L.~{David},
  A.~{Warmflash}, S.~{Murray}, P.E.J. {Nulsen}, \apjl \textbf{567}, L115
  (2002).
\newblock \doi{10.1086/340114}

\bibitem{buot03a}
D.A. {Buote}, A.D. {Lewis}, F.~{Brighenti}, W.G. {Mathews}, \apj \textbf{594},
  741 (2003)

\bibitem{gast09a}
F.~{Gastaldello}, D.A. {Buote}, P.~{Temi}, F.~{Brighenti}, W.G. {Mathews},
  S.~{Ettori}, \apj \textbf{693}, 43 (2009).
\newblock \doi{10.1088/0004-637X/693/1/43}

\bibitem{davi09a}
L.P. {David}, C.~{Jones}, W.~{Forman}, P.~{Nulsen}, J.~{Vrtilek},
  E.~{O'Sullivan}, S.~{Giacintucci}, S.~{Raychaudhury}, \apj \textbf{705}, 624
  (2009).
\newblock \doi{10.1088/0004-637X/705/1/624}

\bibitem{davi94}
L.P. {David}, C.~{Jones}, W.~{Forman}, S.~{Daines}, \apj \textbf{428}, 544
  (1994).
\newblock \doi{10.1086/174264}

\bibitem{buot04b}
D.A. {Buote}, F.~{Brighenti}, W.G. {Mathews}, \apjl \textbf{607}, L91 (2004)

\bibitem{buot02a}
D.A. {Buote}, \apjl \textbf{574}, L135 (2002).
\newblock \doi{10.1086/342532}

\bibitem{buot03b}
D.A. {Buote}, A.D. {Lewis}, F.~{Brighenti}, W.G. {Mathews}, \apj \textbf{595},
  151 (2003)

\bibitem{kim04a}
D.~{Kim}, G.~{Fabbiano}, \apj \textbf{613}, 933 (2004).
\newblock \doi{10.1086/423266}

\bibitem{hump06a}
P.J. {Humphrey}, D.A. {Buote}, \apj \textbf{639}, 136 (2006).
\newblock \doi{10.1086/499323}

\bibitem{mats07a}
K.~{Matsushita}, Y.~{Fukazawa}, J.P. {Hughes}, T.~{Kitaguchi}, K.~{Makishima},
  K.~{Nakazawa}, T.~{Ohashi}, N.~{Ota}, T.~{Tamura}, M.~{Tozuka}, T.G. {Tsuru},
  Y.~{Urata}, N.Y. {Yamasaki}, \pasj \textbf{59}, 327 (2007)

\bibitem{rasm07a}
J.~{Rasmussen}, T.J. {Ponman}, \mnras \textbf{380}, 1554 (2007).
\newblock \doi{10.1111/j.1365-2966.2007.12191.x}

\bibitem{komi09a}
M.~{Komiyama}, K.~{Sato}, R.~{Nagino}, T.~{Ohashi}, K.~{Matsushita}, \pasj
  \textbf{61}, 337 (2009)

\bibitem{deproj}
A.C. {Fabian}, E.M. {Hu}, L.L. {Cowie}, J.~{Grindlay}, \apj \textbf{248}, 47
  (1981).
\newblock \doi{10.1086/159128}

\bibitem{kris83}
G.A. {Kriss}, D.F. {Cioffi}, C.R. {Canizares}, \apj \textbf{272}, 439 (1983).
\newblock \doi{10.1086/161311}

\bibitem{nuls95a}
P.E.J. {Nulsen}, H.~{Bohringer}, \mnras \textbf{274}, 1093 (1995)

\bibitem{buot00c}
D.A. {Buote}, \apj \textbf{539}, 172 (2000).
\newblock \doi{10.1086/309224}

\bibitem{math78a}
W.G. {Mathews}, \apj \textbf{219}, 413 (1978).
\newblock \doi{10.1086/155794}

\bibitem{fabr80a}
D.~{Fabricant}, M.~{Lecar}, P.~{Gorenstein}, \apj \textbf{241}, 552 (1980).
\newblock \doi{10.1086/158369}

\bibitem{davi01a}
L.P. {David}, P.E.J. {Nulsen}, B.R. {McNamara}, W.~{Forman}, C.~{Jones},
  T.~{Ponman}, B.~{Robertson}, M.~{Wise}, \apj \textbf{557}, 546 (2001).
\newblock \doi{10.1086/322250}

\bibitem{voig06a}
L.M. {Voigt}, A.C. {Fabian}, \mnras \textbf{368}, 518 (2006).
\newblock \doi{10.1111/j.1365-2966.2006.10199.x}

\bibitem{nuls10a}
P.E.J. {Nulsen}, S.L. {Powell}, A.~{Vikhlinin}, ArXiv e-prints  (2010)

\bibitem{das10a}
P.~{Das}, O.~{Gerhard}, E.~{Churazov}, I.~{Zhuravleva}, ArXiv e-prints  (2010)

\bibitem{merr94a}
D.~{Merritt}, B.~{Tremblay}, \aj \textbf{108}, 514 (1994).
\newblock \doi{10.1086/117088}

\bibitem{mago99a}
J.~{Magorrian}, \mnras \textbf{302}, 530 (1999).
\newblock \doi{10.1046/j.1365-8711.1999.02135.x}

\bibitem{cava76a}
A.~{Cavaliere}, R.~{Fusco-Femiano}, \aap \textbf{49}, 137 (1976)

\bibitem{sara77a}
C.L. {Sarazin}, J.N. {Bahcall}, \apjs \textbf{34}, 451 (1977).
\newblock \doi{10.1086/190457}

\bibitem{beta}
A.~{Cavaliere}, R.~{Fusco-Femiano}, \aap \textbf{70}, 677 (1978)

\bibitem{daw97a}
D.A. {White}, C.~{Jones}, W.~{Forman}, \mnras \textbf{292}, 419 (1997)

\bibitem{alle01d}
S.W. {Allen}, S.~{Ettori}, A.C. {Fabian}, \mnras \textbf{324}, 877 (2001).
\newblock \doi{10.1046/j.1365-8711.2001.04318.x}

\bibitem{cava09a}
A.~{Cavaliere}, A.~{Lapi}, R.~{Fusco-Femiano}, \apj \textbf{698}, 580 (2009).
\newblock \doi{10.1088/0004-637X/698/1/580}

\bibitem{fusc09a}
R.~{Fusco-Femiano}, A.~{Cavaliere}, A.~{Lapi}, \apj \textbf{705}, 1019 (2009).
\newblock \doi{10.1088/0004-637X/705/1/1019}

\bibitem{fabi86a}
A.C. {Fabian}, P.A. {Thomas}, R.E. {White}, III, S.M. {Fall}, \mnras
  \textbf{221}, 1049 (1986)

\bibitem{balb01a}
S.A. {Balbus}, \apj \textbf{562}, 909 (2001).
\newblock \doi{10.1086/323875}

\bibitem{quat08a}
E.~{Quataert}, \apj \textbf{673}, 758 (2008).
\newblock \doi{10.1086/525248}

\bibitem{parr10a}
I.J. {Parrish}, E.~{Quataert}, P.~{Sharma}, \apjl \textbf{712}, L194 (2010).
\newblock \doi{10.1088/2041-8205/712/2/L194}

\bibitem{rusz10a}
M.~{Ruszkowski}, S.P. {Oh}, \apj \textbf{713}, 1332 (2010).
\newblock \doi{10.1088/0004-637X/713/2/1332}

\bibitem{tozz01a}
P.~{Tozzi}, C.~{Norman}, \apj \textbf{546}, 63 (2001).
\newblock \doi{10.1086/318237}

\bibitem{voit05a}
G.M. {Voit}, S.T. {Kay}, G.L. {Bryan}, \mnras \textbf{364}, 909 (2005).
\newblock \doi{10.1111/j.1365-2966.2005.09621.x}

\bibitem{osul04a}
E.~{O'Sullivan}, T.J. {Ponman}, \mnras \textbf{354}, 935 (2004).
\newblock \doi{10.1111/j.1365-2966.2004.08257.x}

\bibitem{osul07a}
E.~{O'Sullivan}, A.J.R. {Sanderson}, T.J. {Ponman}, \mnras \textbf{380}, 1409
  (2007).
\newblock \doi{10.1111/j.1365-2966.2007.12229.x}

\bibitem{nagi09a}
R.~{Nagino}, K.~{Matsushita}, \aap \textbf{501}, 157 (2009).
\newblock \doi{10.1051/0004-6361/200810978}

\bibitem{buot96d}
D.A. {Buote}, C.R. {Canizares}, \apj \textbf{468}, 184 (1996).
\newblock \doi{10.1086/177680}

\bibitem{buot97a}
D.A. {Buote}, C.R. {Canizares}, \apj \textbf{474}, 650 (1997).
\newblock \doi{10.1086/303490}

\bibitem{buot02b}
D.A. {Buote}, T.E. {Jeltema}, C.R. {Canizares}, G.P. {Garmire}, \apj
  \textbf{577}, 183 (2002).
\newblock \doi{10.1086/342158}

\bibitem{jone97a}
C.~{Jones}, C.~{Stern}, W.~{Forman}, J.~{Breen}, L.~{David}, W.~{Tucker},
  M.~{Franx}, \apj \textbf{482}, 143 (1997).
\newblock \doi{10.1086/304104}

\bibitem{irwi96a}
J.A. {Irwin}, C.L. {Sarazin}, \apj \textbf{471}, 683 (1996)

\bibitem{brig97a}
F.~{Brighenti}, W.G. {Mathews}, \apjl \textbf{486}, L83 (1997)

\bibitem{sun03a}
M.~{Sun}, W.~{Forman}, A.~{Vikhlinin}, A.~{Hornstrup}, C.~{Jones}, S.S.
  {Murray}, \apj \textbf{598}, 250 (2003).
\newblock \doi{10.1086/378887}

\bibitem{osul07b}
E.~{O'Sullivan}, J.M. {Vrtilek}, D.E. {Harris}, T.J. {Ponman}, \apj
  \textbf{658}, 299 (2007).
\newblock \doi{10.1086/511778}

\bibitem{zhan07a}
Z.~{Zhang}, H.~{Xu}, Y.~{Wang}, T.~{An}, Y.~{Xu}, X.~{Wu}, \apj \textbf{656},
  805 (2007).
\newblock \doi{10.1086/510281}

\bibitem{hoek05a}
H.~{Hoekstra}, B.C. {Hsieh}, H.K.C. {Yee}, H.~{Lin}, M.D. {Gladders}, \apj
  \textbf{635}, 73 (2005).
\newblock \doi{10.1086/496913}

\bibitem{klein06a}
M.~{Kleinheinrich}, P.~{Schneider}, H.~{Rix}, T.~{Erben}, C.~{Wolf},
  M.~{Schirmer}, K.~{Meisenheimer}, A.~{Borch}, S.~{Dye}, Z.~{Kovacs},
  L.~{Wisotzki}, \aap \textbf{455}, 441 (2006).
\newblock \doi{10.1051/0004-6361:20042606}

\bibitem{mand06a}
R.~{Mandelbaum}, U.~{Seljak}, G.~{Kauffmann}, C.M. {Hirata}, J.~{Brinkmann},
  \mnras \textbf{368}, 715 (2006).
\newblock \doi{10.1111/j.1365-2966.2006.10156.x}

\bibitem{heym06a}
C.~{Heymans}, E.F. {Bell}, H.~{Rix}, M.~{Barden}, A.~{Borch}, J.A.R.
  {Caldwell}, D.H. {McIntosh}, K.~{Meisenheimer}, C.Y. {Peng}, C.~{Wolf},
  S.V.W. {Beckwith}, B.~{H{\"a}u{\ss}ler}, K.~{Jahnke}, S.~{Jogee}, S.F.
  {S{\'a}nchez}, R.~{Somerville}, L.~{Wisotzki}, \mnras \textbf{371}, L60
  (2006).
\newblock \doi{10.1111/j.1745-3933.2006.00208.x}

\bibitem{prad03a}
F.~{Prada}, M.~{Vitvitska}, A.~{Klypin}, J.A. {Holtzman}, D.J. {Schlegel}, E.K.
  {Grebel}, H.~{Rix}, J.~{Brinkmann}, T.A. {McKay}, I.~{Csabai}, \apj
  \textbf{598}, 260 (2003).
\newblock \doi{10.1086/378669}

\bibitem{shen10a}
J.~{Shen}, K.~{Gebhardt}, \apj \textbf{711}, 484 (2010).
\newblock \doi{10.1088/0004-637X/711/1/484}

\bibitem{brya98a}
G.L. {Bryan}, M.L. {Norman}, \apj \textbf{495}, 80 (1998).
\newblock \doi{10.1086/305262}

\bibitem{mamo05a}
G.A. {Mamon}, E.L. {{\L}okas}, \mnras \textbf{362}, 95 (2005).
\newblock \doi{10.1111/j.1365-2966.2005.09225.x}

\bibitem{bull01a}
J.S. {Bullock}, T.S. {Kolatt}, Y.~{Sigad}, R.S. {Somerville}, A.V. {Kravtsov},
  A.A. {Klypin}, J.R. {Primack}, A.~{Dekel}, \mnras \textbf{321}, 559 (2001).
\newblock \doi{10.1046/j.1365-8711.2001.04068.x}

\bibitem{macc08a}
A.V. {Macci{\`o}}, A.A. {Dutton}, F.C. {van den Bosch}, \mnras \textbf{391},
  1940 (2008).
\newblock \doi{10.1111/j.1365-2966.2008.14029.x}

\bibitem{sato00a}
S.~{Sato}, F.~{Akimoto}, A.~{Furuzawa}, Y.~{Tawara}, M.~{Watanabe}, Y.~{Kumai},
  \apjl \textbf{537}, L73 (2000).
\newblock \doi{10.1086/312772}

\bibitem{wu00a}
X.~{Wu}, Y.~{Xue}, \apjl \textbf{529}, L5 (2000).
\newblock \doi{10.1086/312451}

\bibitem{khos04a}
H.G. {Khosroshahi}, L.R. {Jones}, T.J. {Ponman}, \mnras \textbf{349}, 1240
  (2004).
\newblock \doi{10.1111/j.1365-2966.2004.07575.x}

\bibitem{buot07a}
D.A. {Buote}, F.~{Gastaldello}, P.J. {Humphrey}, L.~{Zappacosta}, J.S.
  {Bullock}, F.~{Brighenti}, W.G. {Mathews}, \apj \textbf{664}, 123 (2007).
\newblock \doi{10.1086/518684}

\bibitem{gned04a}
O.Y. {Gnedin}, A.V. {Kravtsov}, A.A. {Klypin}, D.~{Nagai}, \apj \textbf{616},
  16 (2004)

\bibitem{chur10a}
E.~{Churazov}, S.~{Tremaine}, W.~{Forman}, O.~{Gerhard}, P.~{Das},
  A.~{Vikhlinin}, C.~{Jones}, H.~{B{\"o}hringer}, K.~{Gebhardt}, \mnras
  \textbf{404}, 1165 (2010).
\newblock \doi{10.1111/j.1365-2966.2010.16377.x}

\bibitem{hump10a}
P.J. {Humphrey}, D.A. {Buote}, \mnras \textbf{403}, 2143 (2010).
\newblock \doi{10.1111/j.1365-2966.2010.16257.x}

\bibitem{laba08a}
F.~{La Barbera}, G.~{Busarello}, P.~{Merluzzi}, I.G. {de la Rosa},
  G.~{Coppola}, C.P. {Haines}, \apj \textbf{689}, 913 (2008).
\newblock \doi{10.1086/592769}

\bibitem{auge10a}
M.W. {Auger}, T.~{Treu}, A.S. {Bolton}, R.~{Gavazzi}, L.V.E. {Koopmans}, P.J.
  {Marshall}, L.A. {Moustakas}, S.~{Burles}, ArXiv e-prints  (2010)

\bibitem{krou01a}
P.~{Kroupa}, \mnras \textbf{322}, 231 (2001).
\newblock \doi{10.1046/j.1365-8711.2001.04022.x}

\bibitem{mara05}
C.~{Maraston}, \mnras \textbf{362}, 799 (2005).
\newblock \doi{10.1111/j.1365-2966.2005.09270.x}

\bibitem{fioc97}
M.~{Fioc}, B.~{Rocca-Volmerange}, \aap \textbf{326}, 950 (1997)

\bibitem{bruz03}
G.~{Bruzual}, S.~{Charlot}, \mnras \textbf{344}, 1000 (2003).
\newblock \doi{10.1046/j.1365-8711.2003.06897.x}

\bibitem{vand93a}
R.P. {van der Marel}, M.~{Franx}, \apj \textbf{407}, 525 (1993).
\newblock \doi{10.1086/172534}

\bibitem{gebh00b}
K.~{Gebhardt}, D.~{Richstone}, J.~{Kormendy}, T.R. {Lauer}, E.A. {Ajhar},
  R.~{Bender}, A.~{Dressler}, S.M. {Faber}, C.~{Grillmair}, J.~{Magorrian},
  S.~{Tremaine}, \aj \textbf{119}, 1157 (2000).
\newblock \doi{10.1086/301240}

\bibitem{vand08a}
R.C.E. {van den Bosch}, G.~{van de Ven}, E.K. {Verolme}, M.~{Cappellari}, P.T.
  {de Zeeuw}, \mnras \textbf{385}, 647 (2008).
\newblock \doi{10.1111/j.1365-2966.2008.12874.x}

\bibitem{gebh09a}
K.~{Gebhardt}, J.~{Thomas}, \apj \textbf{700}, 1690 (2009).
\newblock \doi{10.1088/0004-637X/700/2/1690}

\bibitem{schw79a}
M.~{Schwarzschild}, \apj \textbf{232}, 236 (1979).
\newblock \doi{10.1086/157282}

\bibitem{vand98a}
R.P. {van der Marel}, N.~{Cretton}, P.T. {de Zeeuw}, H.~{Rix}, \apj
  \textbf{493}, 613 (1998).
\newblock \doi{10.1086/305147}

\bibitem{rich88a}
D.O. {Richstone}, S.~{Tremaine}, \apj \textbf{327}, 82 (1988).
\newblock \doi{10.1086/166171}

\bibitem{math03c}
W.G. {Mathews}, F.~{Brighenti}, \apj \textbf{599}, 992 (2003).
\newblock \doi{10.1086/379537}

\bibitem{ciot04a}
L.~{Ciotti}, S.~{Pellegrini}, \mnras \textbf{350}, 609 (2004).
\newblock \doi{10.1111/j.1365-2966.2004.07670.x}

\bibitem{binn90a}
J.J. {Binney}, R.L. {Davies}, G.D. {Illingworth}, \apj \textbf{361}, 78 (1990).
\newblock \doi{10.1086/169169}

\bibitem{roma09a}
A.J. {Romanowsky}, J.~{Strader}, L.R. {Spitler}, R.~{Johnson}, J.P. {Brodie},
  D.A. {Forbes}, T.~{Ponman}, \aj \textbf{137}, 4956 (2009).
\newblock \doi{10.1088/0004-6256/137/6/4956}

\bibitem{thom07a}
J.~{Thomas}, R.P. {Saglia}, R.~{Bender}, D.~{Thomas}, K.~{Gebhardt},
  J.~{Magorrian}, E.M. {Corsini}, G.~{Wegner}, \mnras \textbf{382}, 657 (2007).
\newblock \doi{10.1111/j.1365-2966.2007.12434.x}

\bibitem{rusl10a}
S.P. {Rusli}, J.~{Thomas}, P.~{Erwin}, R.P. {Saglia}, N.~{Nowak}, R.~{Bender},
  ArXiv e-prints  (2010)

\end{thebibliography}

\end{document}